\documentclass[11pt]{article}

\usepackage{algorithm}
\usepackage{algpseudocode}
\usepackage{amsfonts}
\usepackage{amsmath}
\usepackage{amsthm}
\usepackage{booktabs}
\usepackage[left=25mm,right=25mm,top=15mm,bottom=18mm]{geometry}
\usepackage{graphicx}
\usepackage{hyperref}
\usepackage{multirow}
\usepackage{subcaption}
\usepackage{times}
\usepackage{url}
\usepackage{xspace}

\def\hcf{HC-Full\xspace}
\def\hca{HC-Sep\xspace}
\def\hcb{HC-Acc\xspace}
\def\ihc{IHC\xspace}
\def\ihcre{IHC-Re\xspace}

\def\ihci#1{IHC-iKL0.{#1}}
\def\ihcw#1{IHC-wKL0.{#1}}

\def\ihcrei#1{IHC-Re-iKL0.{#1}}
\def\ihcrew#1{IHC-Re-wKL0.{#1}}

\title{Batchwise Probabilistic Incremental Data Cleaning}
\date{November 2020}

\author{\makebox[0.40\linewidth]{Paulo H. Oliveira\footnote{Both authors contributed equally to this research.}} \\
        pholiveira@usp.br \\
        University of S\~{a}o Paulo \\
        S\~{a}o Carlos, SP, Brazil
   \and \makebox[0.40\linewidth]{Daniel S. Kaster\footnotemark[1]} \\
        dskaster@uel.br \\
        University of Londrina \\
        Londrina, PR, Brazil
   \and \makebox[0.40\linewidth]{Caetano Traina-Jr.} \\
        caetano@icmc.usp.br \\
        University of S\~{a}o Paulo \\
        S\~{a}o Carlos, SP, Brazil
   \and \makebox[0.40\linewidth]{Ihab F. Ilyas} \\
        ilyas@uwaterloo.ca \\
        University of Waterloo \\
        Waterloo, ON, Canada
}

\begin{document}
\maketitle

\begin{abstract}
Lack of data and data quality issues are among the main bottlenecks that prevent further artificial intelligence adoption within many organizations, pushing data scientists to spend most of their time cleaning data before being able to answer analytical questions.
Hence, there is a need for more effective and efficient data cleaning solutions, which, not surprisingly, is rife with theoretical and engineering problems.
This report addresses the problem of performing holistic data cleaning incrementally, given a fixed rule set and an evolving categorical relational dataset acquired in sequential batches.
To the best of our knowledge, our contributions compose the first incremental framework that cleans data \textbf{(i)} independently of user interventions, \textbf{(ii)} without requiring knowledge about the incoming dataset, such as the number of classes per attribute, and \textbf{(iii)} holistically, enabling multiple error types to be repaired simultaneously, and thus avoiding conflicting repairs.
Extensive experiments show that our approach outperforms the competitors with respect to repair quality, execution time, and memory consumption.
\end{abstract}

\section{Introduction}

Enterprises have been collecting large amounts of data from a variety of sources to power their applications, seeking richer and better-informed analytics \cite{Ilyas2016Effective}.
Data acquisition often introduces problems in data, such as missing values, typos, replicated entries for the same real-world entity, and data quality constraint violations.
Therefore, there is a need for creating more effective and efficient data cleaning solutions, which, not surprisingly, is rife with theoretical and engineering problems.

Data cleaning consists of two main phases:
\textbf{(i)} Error Detection, where data errors and constraint violations are identified and possibly validated by experts; and
\textbf{(ii)} Error Repairing, where updates are made (or suggested) to bring the data to a cleaner state \cite{Ilyas2019Data}.
New records may arrive constantly, and data quality constraints might also evolve over time.
As new data are continuously collected and included to the existing, previously cleaned data, the process of cleaning each new snapshot of data is challenging and time-consuming, thus it is impractical to re-execute the whole data cleaning pipeline every time that data or constraints change.
Therefore, performing data cleaning incrementally is paramount in the context of evolving data or constraints \cite{Ilyas2015Trends}.
Incremental data cleaning is even more important when the data do not fit into main memory, which tends to be the case for data profiling tasks \cite{Naumann2014Data} such as constraint discovery and error detection.

With the increasing availability of resources for building large-scale Machine Learning solutions, employing ML techniques for data cleaning has become a promising direction.
For instance, some works have been modelling the task of repairing categorical data as a multi-class classification problem \cite{Volkovs2014Continuous,Rekatsinas2017HoloClean}.
In this line of work, ML models are built using a probabilistic approach on the data, which brings the advantage of enabling a \textit{holistic} treatment over a variety of data errors, such as typos and constraint violations, rather than tackling each problem in isolation \cite{Ilyas2019Data}.
Multiple types of features can be combined to feed ML models, such as features based on quantitative statistics \cite{Beskales2013OnTheRelative,Papotti2013Holistic} and integrity constraints \cite{Mayfield2010ERACER,Yakout2013SCAREd}.
The approach of treating various error types and data features holistically is a recent trend \cite{Papotti2013Holistic,Abedjan2016Detecting,Rekatsinas2017HoloClean,Heidari2019HoloDetect} in the data cleaning literature.

Some state-of-the-art techniques can perform data cleaning holistically, but not incrementally \cite{Papotti2013Holistic,Rekatsinas2017HoloClean,Heidari2019HoloDetect}.
Conversely, current incremental data cleaning approaches either
\textbf{(i)} present limitations, such as dealing with a single error type and/or depending on user interactions \cite{Volkovs2014Continuous}, or
\textbf{(ii)} perform data cleaning targeting a better ML model just for the task and dataset at hand, rather than focusing on the cleaned dataset itself \cite{Krishnan2016ActiveClean}.
Hence, the literature lacks an ML-based data cleaning solution that is both holistic and incremental.

In ML-based data cleaning solutions, large evolving datasets are continuously repaired as new parts of the dataset arrive --- throughout this report, we refer to such a new part as a \textbf{batch} or \textbf{data batch}.
The ML model is eventually updated according to data changes.
This series of tasks calls for an end-to-end ML management pipeline, which is a line of work that has been drawing increasing attention.
Major players such as Google and Amazon have been working to offer comprehensive frameworks for deploying ML systems \cite{Baylor2017TFX,Schelter2018OnChallenges}.
Such frameworks aim at supporting all the life cycle of an ML application, including different aspects of model management.
These frameworks represent an essential first step towards managing ML applications, but they are general-purpose.
To perform end-to-end incremental data cleaning, a new solution is needed.

\subsection*{Technical Challenges}

In an incremental scenario for holistic data cleaning, the following challenges arise.
\begin{itemize}
    \item \textbf{How to select data cells to inspect for errors.}
    When looking for constraint violations, we must inspect cells from distinct tuples to identify patterns that should not occur.
    It is possible that certain cells within existing data do not violate any constraint in the beginning, but reveal themselves as erroneous as new tuples arrive and new patterns show up.
    In this context, we come across two possible measures:
    \textbf{(i)} take into account the whole data seen so far when spotting errors to be repaired, which can be increasingly costly as new batches arrive, or
    \textbf{(ii)} assume that few errors (if any) will occur within the existing data when more tuples arrive, thus looking for errors only across the incoming tuples, which is expected to be less costly than inspecting the whole data.

    \item \textbf{How to incrementally generate features.}
    Features are measurable properties of a phenomenon being observed.
    In an evolving dataset, it is natural to expect evolving features as well.
    Instead of computing features from scratch whenever a new data batch arrives, an incremental data cleaning system should be able to generate features incrementally, allowing this step to be less time-consuming.
    For instance, if the features are based on quantitative statistics, these should be updated whenever a new batch arrives, preferably without recomputing the statistics from the very beginning.

    \item \textbf{Whether to retrain ML models as new batches arrive.}
    The ability to identify significant changes in the incoming data allows model training optimizations.
    For example, if an incoming data batch follows the same (or a similar) distribution of the training data, it is expected that the model will improve little when retrained with examples from the new batch.
    Hence, an incremental approach should be able to automatically trigger retraining only when needed, which calls for proper metrics to guide the process.
\end{itemize}

\subsection*{Contributions}

This report describes our end-to-end holistic incremental data cleaning framework.
To the best of our knowledge, this is the first approach to perform incremental data cleaning holistically (that is, combining multiple error detection criteria) and independently of user interactions.
Specifically, our contributions are as follows:
\begin{enumerate}
    \item We extend the framework used by HoloClean\footnote{\url{http://www.holoclean.io/}}, a state-of-the-art holistic data cleaning system \cite{Rekatsinas2017HoloClean,Roh2019Survey}, enhancing its existing modules and including a new one to enable incremental cleaning (Section \ref{sec:framework_overview}).

    \item Based on the existing implementation of HoloClean, we instantiate the proposed framework by developing a system for incrementally cleaning structured categorical data.
    Initially, we describe our system from the data loading to the feature generation steps.
    Specifically, we detail how we
    \textbf{(i)} identify errors,
    \textbf{(ii)} compute dataset statistics, and
    \textbf{(iii)} use error information and statistics to generate the data features that feed our ML models for training.
    We also present our strategy for skipping model retraining based on data distribution changes.
    It allows skipping retraining when an incoming batch does not bring significant changes in the data distribution, which occurs as early as before computing dataset statistics, thus avoiding needless feature generation and enabling immediate data repair inference (Section \ref{sec:incremental_features}).

    \item We further describe the system focusing on the steps after feature generation.
    We detail our approach for setting up the ML models, which differs from the one used by HoloClean.
    Distinctly from the one-model setup in the original version of HoloClean, ours uses one model per attribute in the dataset.
    This allows reducing memory requirements and avoids unnecessary model retraining more often, since the models are separate (Section \ref{sec:model_training}).

    \item We extensively evaluate our approach against competitors over publicly available categorical datasets with varying cardinalities, measuring repair quality based on how clean is the resulting dataset, as well as execution time and memory consumption (Section \ref{sec:experiments}).
\end{enumerate}

\subsection*{Problem Statement}

Let $D$ be a relation with the schema $(A_{1}, \ldots, A_{N})$, where $A_{i}$ is an attribute and $N$ is the dimensionality of $D$.
Furthermore, let $D\{t_{1}, \ldots, t_{M}\}$ be the set of tuples in $D$.
For every tuple $t_{j} \in D$, each attribute stores a single value, and $M$ is the cardinality of $D$.
We refer to an attribute value in a tuple $t[A_{i}]$ as a cell.
A cell $c$ is dirty, i.e. it is an error, if its observed value $v_{c}$ is different from its unknown true value $v_{c}^{*}$.
All dirty cells in the relation $D$ correspond to the erroneous subset $D^{E}$, whereas the clean subset $D^{C} = D \setminus D^{E}$ comprises the clean cells in $D$.
A cell repair is an attribution of a value $\hat{v}_{c}$ to the cell, where $\hat{v}_{c} \neq v_{c}$, and the repair is correct if $\hat{v}_{c} = v_{c}^{*}$.

Suppose $D$ is an evolving dataset with batches $D_{1}, D_{2}, \dots, D_{B}$, such that $D = \bigcup_{k=1}^B D_{k}$.
At a certain point in time $k$, only the batches $D_{1..k}$ are known.
The cardinality $|D_{k}|$ of each batch can be distinct.
The dirty cells in a batch $D_{k}$ correspond to the erroneous subset $D_{k}^{E}$ of that batch, and the clean subset $D_{k}^{C} = D_{k} \setminus D_{k}^{E}$ comprises the clean cells of such batch.

Finally, let $\Lambda = (\mathcal{C}_{e}, \mathcal{C}_{f}, \mathcal{C}_{t})$ be an incremental data cleaning process, where $\mathcal{C}_{e}$ is the set of criteria for error detection, $\mathcal{C}_{f}$ is the set of criteria for feature vector generation, and $\mathcal{C}_{t}$ is the set of criteria for ML model training.
Moreover, let $\hat{D}_{1..k}^{E} \subseteq D_{1..k}^{E}$ be the set of dirty cells identified according to $\mathcal{C}_{e}$.
Given $\Lambda$ and the set of batches $D_{1..k}$ of an evolving dataset $D$ without ground-truth readily available, \textit{the problem addressed in this report is to infer a repair for each dirty cell $c \in \hat{D}_{1..k}^{E}$, employing features generated based on $\mathcal{C}_{f}$ and models trained based on $\mathcal{C}_{t}$}.

\section{Preliminaries}
\label{sec:preliminaries}

\subsection{Data Cleaning Pipeline}

We describe the basic steps that compose the execution of a data cleaning system.
Variations can be applied depending on the requirements, such as performing multiple iterations of the cleaning process and including user validation at the end of each iteration.
Here, we focus on the basic steps that underlie our framework.

\textbf{Data Loading.}
The system loads the raw dataset, which is kept in memory to be further processed and can be stored in a Database Management System (DBMS) as well.

\textbf{Error Detection.}
The system runs error detectors to identify potentially dirty cells, which are values in the dataset spotted as erroneous based on the criteria employed by the detectors.
An example of a straightforward error detector is the Null Detector, which looks for empty cells and marks them as errors.
Another example is the Constraint Violation Detector, which identifies cells violating a given set of constraints, such as Functional Dependencies (FDs) or Denial Constraints (DCs) \cite{Chu2013Discovering}.

\textbf{Error Repairing.}
This step aims at fixing the newly-found dirty cells in a principled way.
One of the existing operational principles for performing repairs is the \textit{principle of minimality}, which consists of updating a minimal number of cells needed to bring the dataset to a clean state, i.e. free of every spotted error such as wrong values and constraint violations \cite{Kolahi2009OnApproximating}.
However, minimal repairs might not lead to correct results, as the updated cells can still have wrong values even if they are no longer spotted by the error detectors being used \cite{Ilyas2019Data}.
Another operational principle aims for the \textit{most probable value}.
For instance, if a dirty cell belongs to a categorical attribute, this principle can be employed in a multi-class classification problem, in order to infer the correct value as the most probable value from the cell domain.
Conversely, if a dirty cell belongs to a numerical attribute, this principle can be employed in a regression problem, in order to learn a model from the data (the dataset itself and possibly external data) to predict the correct value.
Our approach employs the \textit{most probable value} principle.

\subsection{The HoloClean System}

HoloClean is a state-of-the-art data cleaning system \cite{Roh2019Survey} that combines multiple signals, such as quantitative statistics, quality rules, and reference data, to infer data repairs.
The original version of the system \cite{Rekatsinas2017HoloClean} builds a Probabilistic Graphical Model (PGM) \cite{Koller2009Probabilistic} whose random variables capture the uncertainty over cells in the input dataset.
Specifically, given a dataset $D$, the system associates each cell $c \in D$ with a random variable $T_{c}$ that takes values from a finite domain $dom(c)$ and compiles a PGM that describes the data distribution of random variables $T_{c}$.

The PGM created by HoloClean is also known as a \textit{factor graph}.
A factor graph is a hypergraph $(T, F, \theta)$, where $T$ is a set of nodes corresponding to random variables and $F$ is a set of hyperedges.
Each hyperedge $\phi \in F$ can be referred to as a factor, where $\phi \subseteq T$, and is associated with a factor function $h_{\phi}$ and a real-valued weight $\theta_{\phi}$.
The factor function takes an assignment of the random variables in $\phi$ and returns a value in $\{-1, 1\}$, that is, $h_{\phi} : \mathbb{D}^{|\phi|} \rightarrow \{-1, 1\}$.
Hyperedges $\phi$, functions $h_{\phi}$, and weights $\theta_{\phi}$ define a factorization of the probability distribution $P(T)$ as:
\begin{equation}
    P(T) = \dfrac{1}{Z} \exp \left( \sum_{\phi \in F} \theta_{\phi} \cdot h_{\phi}(\phi) \right),
\end{equation}
\noindent
where $Z$ is a partition function, which is a normalization constant ensuring the resulting distribution is valid.

A set of error detectors can be used to identify uncertain cells in $D$, e.g. the aforementioned Null Detector and Constraint Violation Detector.
The error detection phase splits the random variables $T_{c}$ into sets $T_{c}^{u}$ and $T_{c}^{k}$, where $T_{c}^{u}$ is the set of random variables referring to uncertain cells (whose values need to be inferred) and $T_{c}^{k}$ is the set of random variables referring to clean cells (whose observed values are used for training).
Error detection criteria can be encoded as factors in the factor graph, along with other signals such as quantitative statistics.
The first version of HoloClean employs the DeepDive system \cite{Shin2015Incremental} to declaratively specify random variables and compile various signals as factors.

After the factor graph is built, its parameters (factor weights) are learned in a weakly supervised fashion: HoloClean uses the set of cells identified as clean to generate a large number of correct and incorrect labelled examples of random variables in $T_{c}^{k}$, where the current observed values correspond to the ``correct'' examples and multiple generated values from the domain correspond to a set of ``incorrect'' examples.
Then, assuming $T$ to be the set of all random variables $T_{c}$, HoloClean employs Empirical Risk Minimization (ERM) over the log-likelihood $P(T)$ to estimate the parameters of its factor graph.
Finally, the system estimates the value $\hat{v}_{c}$ of each uncertain cell in $T_{c}^{u}$ by performing approximate inference via Gibbs sampling \cite{Zhang2014DimmWitted}, assigning those cells to the corresponding Maximum A Posteriori (MAP) estimates of variables $T_{c}^{u}$.

\section{Framework Overview}
\label{sec:framework_overview}

\begin{figure}[!ht]
    \centering
    \includegraphics[width=0.70\textwidth]{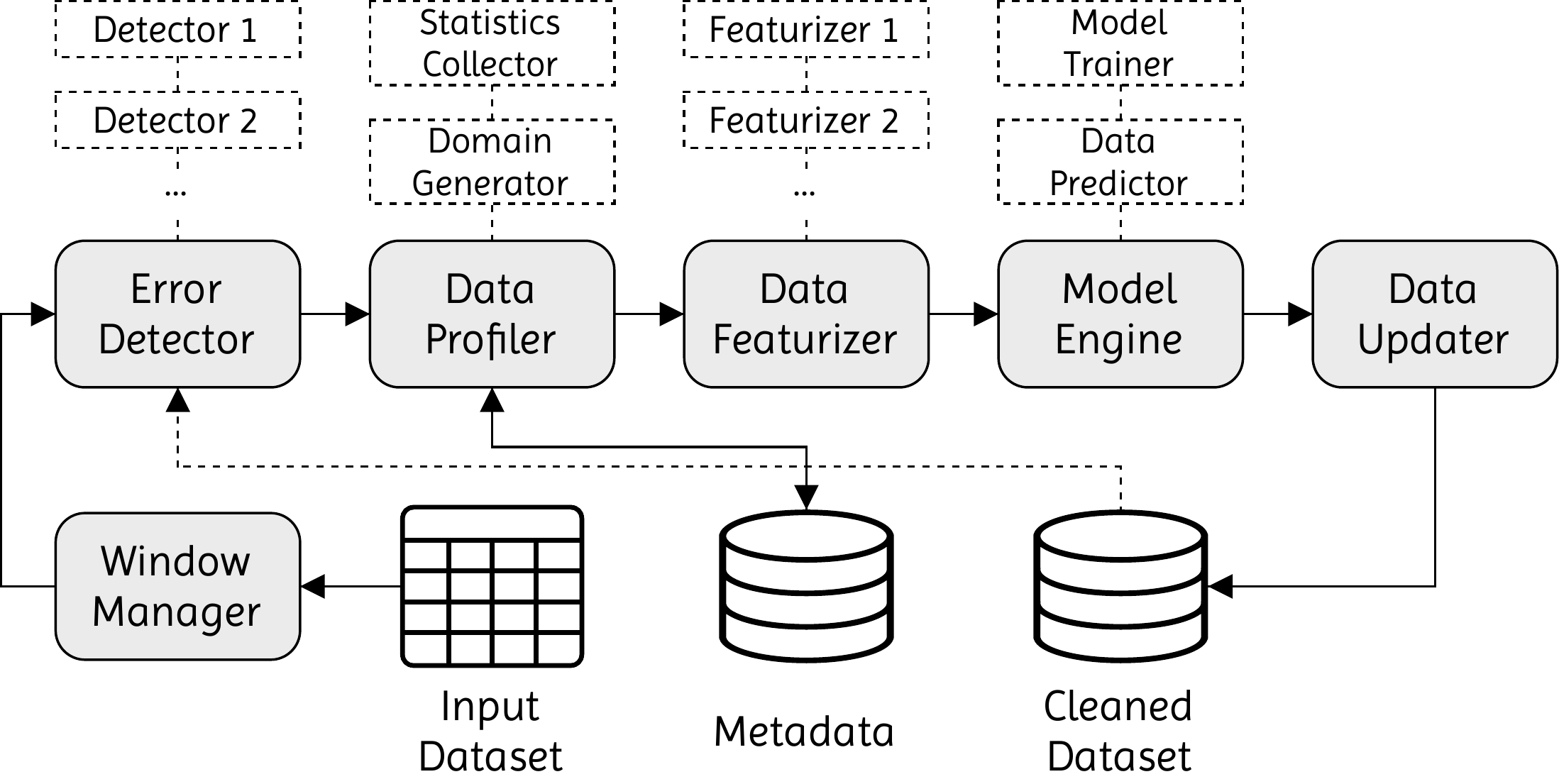}
    \caption{Architecture of the proposed framework.}
    \label{fig:architecture}
\end{figure}

We describe our framework following the architecture depicted in Figure \ref{fig:architecture}, which extends the architecture on which the HoloClean system is based.
The \textit{Window Manager} is a brand new module, whereas the remaining modules are already part of the HoloClean system, but which have also been enhanced for our framework.

The \textit{Window Manager} is responsible for handling the incoming \textit{Input Dataset}, generating the data windows (or batches) to be processed.
Data batches can be defined according to memory resources or time limits, and the number of elements within a data window can vary over time.
The window size affects
\textbf{(i)} how much new cleaned data is produced by the cleaning process, as well as
\textbf{(ii)} the extent to which the existing dataset can be enriched upon receiving new cleaned data, which can improve ML model training.

The \textit{Error Detector} module receives each data window from the \textit{Window Manager} and looks for potentially dirty cells.
It applies one or more detectors, such as for identifying data cells whose values are missing or that violate integrity constraints.
Errors can be detected either solely within the data window itself or involving cells both in the data window and in the clean subset (if any) of the dataset.
The parts of the dataset to be inspected for errors are determined by the criteria $\mathcal{C}_{e}$ in $\Lambda$.

The \textit{Data Profiler} initially employs the Statistics Collector to gather statistics from the data.
The statistics can account either for all the dataset seen so far or for a subset of it, as determined by the set of criteria $\mathcal{C}_{f}$.
Then, based on the collected statistics, the \textit{Data Profiler} executes the Domain Generator to compute $dom(c)$ for every cell $c$ to be used either in model training or in data repairing, where $dom(c)$ is the set of all values that can be used as evidence for repairing cell $c$.
The set of criteria $\mathcal{C}_{f}$ is also responsible for defining what statistics must be computed.
The statistics are stored as part of the \textit{Metadata} in an underlying DBMS and are incrementally updated as the dataset evolves.

The \textit{Data Featurizer} relies on one or multiple Featurizers to generate features from the cells to be used in model training and data repairing.
Examples of features include attribute-value frequencies, co-occurrences of attribute-value pairs, and constraint violations.
When multiple Featurizers are employed, the \textit{Data Featurizer} module concatenates the features into a composite feature vector for each cell, enabling the aforementioned holistic treatment of multiple signals in model training.

The \textit{Model Manager} module consists of a Model Trainer and a Data Repairer.
The Model Trainer can train one or more models, and the criteria $\mathcal{C}_{t}$ determine the model settings, such as the number of models and what portion of the training data each model is assigned to.
After model training, the Data Repairer predicts the repairs and forwards them to the \textit{Data Updater} module, which persists the repaired data in the DBMS by updating the Cleaned Dataset.
Finally, when the next batch arrives, the Cleaned Dataset will be available to be examined (which can or cannot occur depending on the sets of criteria $\mathcal{C}_{e}$ and $\mathcal{C}_{f}$) by the \textit{Error Detector} and \textit{Data Profiler} modules.

\section{Incremental Features}
\label{sec:incremental_features}

Our proposal performs incremental data cleaning holistically, which consists of taking into account multiple signals to enrich the model training step, allowing us to generate distinct types of data features.
We carried out a preliminary evaluation to analyze the effect of different feature types on repair quality:
\textbf{(i)} frequency count of each cell value across the whole dataset;
\textbf{(ii)} co-occurrence count of each cell value with the values from the remaining attributes in the relation;
\textbf{(iii)} number of constraint violations in which a cell is involved; and
\textbf{(iv)} embedding features, similar to the ones of word2vec models, but focusing on numerical data.
The preliminary evaluation showed that the co-occurrence and embedding features enabled the best results when inferring repairs, whereas the other feature types provided an overall lower benefit on repair quality.
Since our main interest is in categorical data types, we employ only co-occurrence features in this work.

To generate the features, we first compute statistics from the data.
To do so, our approach considers all the data received so far, i.e. the existing data from previous batches plus the incoming data.
The first set of statistics consists of:
\begin{itemize}
    \item \textbf{single-attribute frequencies}, which maintain one dictionary per attribute in the dataset, where each key in the dictionary is a value from the current attribute domain, and such value is associated with its corresponding frequency count.
    Formally, we define the single-attribute frequencies as
    \begin{equation}
        single\_freq = \{ A_{i}, ~ \{ ( v_{ik}, ~ freq(v_{ik}) ) ~ | ~ 1 \leq i \leq N, ~ 1 \leq k \leq |D[A_{i}]| \} \},
    \end{equation}
    where $A_{i}$ is the $i$-th attribute in relation $D$, $N$ is the number of attributes in the schema of $D$, $v_{ik}$ is the $k$-th value in the value set of $A_{i}$, $freq(v_{ik})$ is the frequency count of $v_{ik}$, and $|D[A_{i}]|$ is the number of distinct values in $A_{i}$.

    \item \textbf{pairwise frequencies}, which maintain one dictionary per attribute pair, where each dictionary keeps track of all co-occurrences between domain values in the corresponding pair.
    The dictionaries have the form \texttt{\{first\_value -> \{second\_value -> count\}\}}, where \texttt{first\_value} is a domain value from the first attribute in the pair, \texttt{second\_value} is a domain value from the second attribute in the pair, and \texttt{count} is the number of times both values co-occur across all the data.
    The pairwise statistics are formally defined as
    \begin{equation}
        \begin{split}
            pair\_freq = \{ A_{i}, ~ \{ A_{i'}, ~ \{ v_{ik}, ~ \{ ( v_{i'k'}, ~ freq(v_{ik}, ~ v_{i'k'}) ) ~ | ~ & i \neq i', \\
            & k \neq k', \\
            & 1 \leq i, i' \leq N, \\
            & 1 \leq k \leq |D[A_{i}]|, \\
            & 1 \leq k' \leq |D[A_{i'}]|\} \} \} \},
        \end{split}
    \end{equation}
\end{itemize}
where $A_{i}$ and $A_{i'}$ are distinct attributes in relation $D$, $v_{ik}$ is the $k$-th value in the value set of $A_{i}$, $v_{i'k'}$ is the $k'$-th value in the value set of $A_{i'}$, $N$ is the number of attributes in the schema of $D$, $freq(v_{ik}, ~ v_{i'k'})$ is the frequency count of $v_{ik}$ and $v_{i'k'}$, $|D[A_{i}]|$ is the quantity of distinct values in $A_{i}$, and $|D[A_{i'}]|$ is the quantity of distinct values in $A_{i'}$.
The incremental maintenance of frequencies computes both the single-attribute and pairwise ones from the current data batch, then adds them to the corresponding frequencies from the previous data batches.
The computational cost of maintaining frequencies is proportional to the batch size.

Another set of statistics consists of \textbf{attribute correlations}, based on the normalized conditional entropy of attribute pairs.
The normalized conditional entropy ranges from $0$ to $1$: it equals $0$ when the attributes are strongly correlated, whereas it equals $1$ when the attributes are independent.
Thus, we reverse the normalized conditional entropy value to reflect the attribute correlation.
We compute the correlation of attributes $x$ and $y$, denoted by $corr(x, y)$, as follows:
\begin{equation}
    corr(x, y) = 1 - norm\_cond\_ent(x, y),
\end{equation}
where $norm\_cond\_ent(x, y)$ is the conditional entropy of $x$ and $y$ normalized by the domain size of attribute $x$.
Such normalization occurs by using the number of distinct values in $x$ (obtained from the previously computed single-attribute frequencies) as the log base in the conditional entropy formula (described in Appendix \ref{sec:incremental_conditional_entropy}).
Specific cases that obviate the normalized conditional entropy computation are: \textbf{(i)} when $x = y$, which assigns the correlation value to $1$, and \textbf{(ii)} when the domain size of $x$ is $1$, which sets the correlation value to $0$.
As the conditional entropy is asymmetric, we need to perform pairwise computations involving all pairs of attributes.
In the context of an evolving dataset, to the best of our knowledge, the literature lacks a principled approach for incrementally maintaining the normalized conditional entropy of an attribute pair.
We have elaborated such an incremental approach, which we formally describe in Appendix \ref{sec:incremental_conditional_entropy}.

The feature vector of a cell has the co-occurrence statistics for all its \emph{candidate values}, which are part of the domain of that cell.
Specifically, the candidate values for a cell are those in its domain which co-occur with the attribute values within the tuple to which that cell belongs.
The set of candidate values is determined this way based on the intuition that values within the same tuple share a relationship, and thus are likely to co-occur in other tuples as well.
That is, given a cell $c$ of a tuple $t$, the only useful values to be considered as evidences for a repair are those appearing in other tuple(s) (any tuple $t' \neq t$) alongside attribute values from tuple $t$.
To visualize this process in a concrete situation, suppose the following example depicting four tuples of a relation, focusing on attributes $B$ and $A$.
\begin{verbatim}
B | A
-----
h   b  <- Current tuple, attribute A under analysis (cell storing "b").
h   c     Cell domain is (b, c, e).
i   d
h   e
\end{verbatim}
Consider that we are generating the cell domain for the first tuple (indicated as \emph{Current tuple} in the example) at attribute $A$, i.e. the cell with value $b$.
We can observe that $b$ co-occurs with $h$ in that tuple, which suggests that value $h$ in $B$ might have a relationship with values from attribute $A$.
Therefore, other values in $A$ co-occurring with value $h$ are candidate values for the cell currently storing $b$.
Hence, based on this reasoning, we have $(b, c, e)$ as the cell domain.

The domain of a cell is formally defined as follows.
Let $A_{i}[t] = v_{it}$ be the value of attribute $A_{i}$ in tuple $t$, assigned to a cell denoted by $c$.
Moreover, let $A_{i'}[t] = v_{i't}$ be the value of an attribute $A_{i'} \neq A_{i}$ in tuple $t$, and let $A_{i}[t'] = v_{it'}$ be the value of attribute $A_{i}$ in another tuple $t' \neq t$.
Finally, let $corr(A', A'')$ be the correlation strength of a pair of attributes $A'$ and $A''$, and let $\Omega$ be a given correlation threshold.
Then, the domain of cell $c$, denoted by $dom(c)$, is the following set:
\begin{equation}
    \begin{split}
        dom(c) = \{ v_{it} \} ~ \cup ~ \{ v_{it'} ~ | ~ & \forall A_{i'} \neq A_{i}, ~ \forall t' \neq t, \\
        & \{ v_{i't}, ~ v_{it'} \} \subset t', \\
        & corr(A_{i'}, ~ A_{i}) > \Omega \},
    \end{split}
\end{equation}
\noindent
where $\{ v_{i't}, v_{it'} \}$ denotes the co-occurrence of $v_{i't}$ with $v_{it'}$.

The full feature vector of a cell $c$ is a 2D tensor.
Its first dimension refers to all values in $dom(c)$, whereas its second dimension refers to all attributes in the dataset.
Specifically, each feature is a number representing the co-occurrence ratio of a value in $dom(c)$ with a value $v_{i't}$ (of another attribute $A_{i'}$ in tuple $t$) in the whole dataset received so far.
Algorithm \ref{alg:generate_feature_vector} shows the feature generation procedure in more detail.
Lines \ref{alg:generate_feature_vector:line_2}--\ref{alg:generate_feature_vector:line_5} initialize and assign variables, including a 2D tensor for the feature vector that is initialized completely with zeros.
The first dimension in the tensor accommodates the maximum domain size of cells from the \texttt{cell\_attr} attribute (which varies between attributes, hence our ability to save memory by employing attribute-specific ML models).
The second dimension, in turn, accommodates the total quantity of attributes in the relation.
Then, the algorithm iterates over all attributes (except when the analyzed attribute is the given \texttt{cell\_attr} attribute itself), generating the co-occurrence features for each candidate value in \texttt{cell\_domain}.
Lines \ref{alg:generate_feature_vector:line_10}--\ref{alg:generate_feature_vector:line_16} obtain the frequency values from the previously-computed statistics, then employ such values to compute the \texttt{ratio} assigned to the corresponding tensor position.
Finally, Algorithm \ref{alg:generate_feature_vector} returns the 2D feature vector tensor for the current data cell.

\begin{algorithm}
\caption{Feature vector generation for data cell from \textit{cell\_attr} attribute}
\label{alg:generate_feature_vector}
\begin{algorithmic}[1]
\footnotesize
\Procedure{generate\_feature\_vector}{$cell\_attr, ~ cell\_domain, ~ tuple$}
\State $max\_domain \gets get\_max\_domain(cell\_attr)$
\label{alg:generate_feature_vector:line_2}
  \Comment{Maximum number of distinct values in \textit{cell\_attr}}
\State $all\_attrs \gets get\_attributes()$
  \Comment{List of attributes in relation}
\State $tensor \gets zeros(1, ~ max\_domain, ~ length(all\_attrs))$
  \Comment{Dimensions for tensor initialization}
\State $single\_stats, ~ pair\_stats \gets get\_computed\_statistics()$
\label{alg:generate_feature_vector:line_5}
\For{$attr ~ in ~ all\_attrs$}
  \If{$attr = cell\_attr$}
    \State $continue$
      \Comment{Skip iteration when attributes are the same}
  \EndIf
  \State $value \gets tuple[attr]$
  \label{alg:generate_feature_vector:line_10}
  \State $value\_freq = single\_stats[attr][value]$
    \Comment{Frequency of \textit{value} across all the data received so far}
  \State $co\_list = pair\_stats[attr][cell\_attr][value]$
    \Comment{Values in \textit{cell\_attr} co-occurring with \textit{value} in \textit{attr}}
  \For{$candidate ~ in ~ cell\_domain$}
    \State $co\_freq \gets co\_list.count(candidate)$
      \Comment{Frequency of \textit{candidate} with \textit{value}}
    \State $ratio \gets co\_freq ~ / ~ val\_freq$
    \State $tensor[0][get\_domain\_position(candidate)][get\_attr\_position(attr)] \gets ratio$
    \label{alg:generate_feature_vector:line_16}
  \EndFor
\EndFor
\State \Return $tensor$
\EndProcedure
\normalsize
\end{algorithmic}
\end{algorithm}

\section{Model Training}
\label{sec:model_training}

The predictive ML model of HoloClean is a single-layer neural network.
The model applies an exponential function over the product between the input feature vector and the learned weights, then applies the \emph{softmax} function.
The weights correspond to the importance of each feature to determine the most probable value of a cell to be repaired.
As a single model is employed for all attributes, the feature vector has one feature set per attribute, and the weights for all attributes are jointly learned.

\begin{figure}[!ht]
    \begin{subfigure}[c]{0.49\linewidth}
        \centering
        \includegraphics[height=4cm]{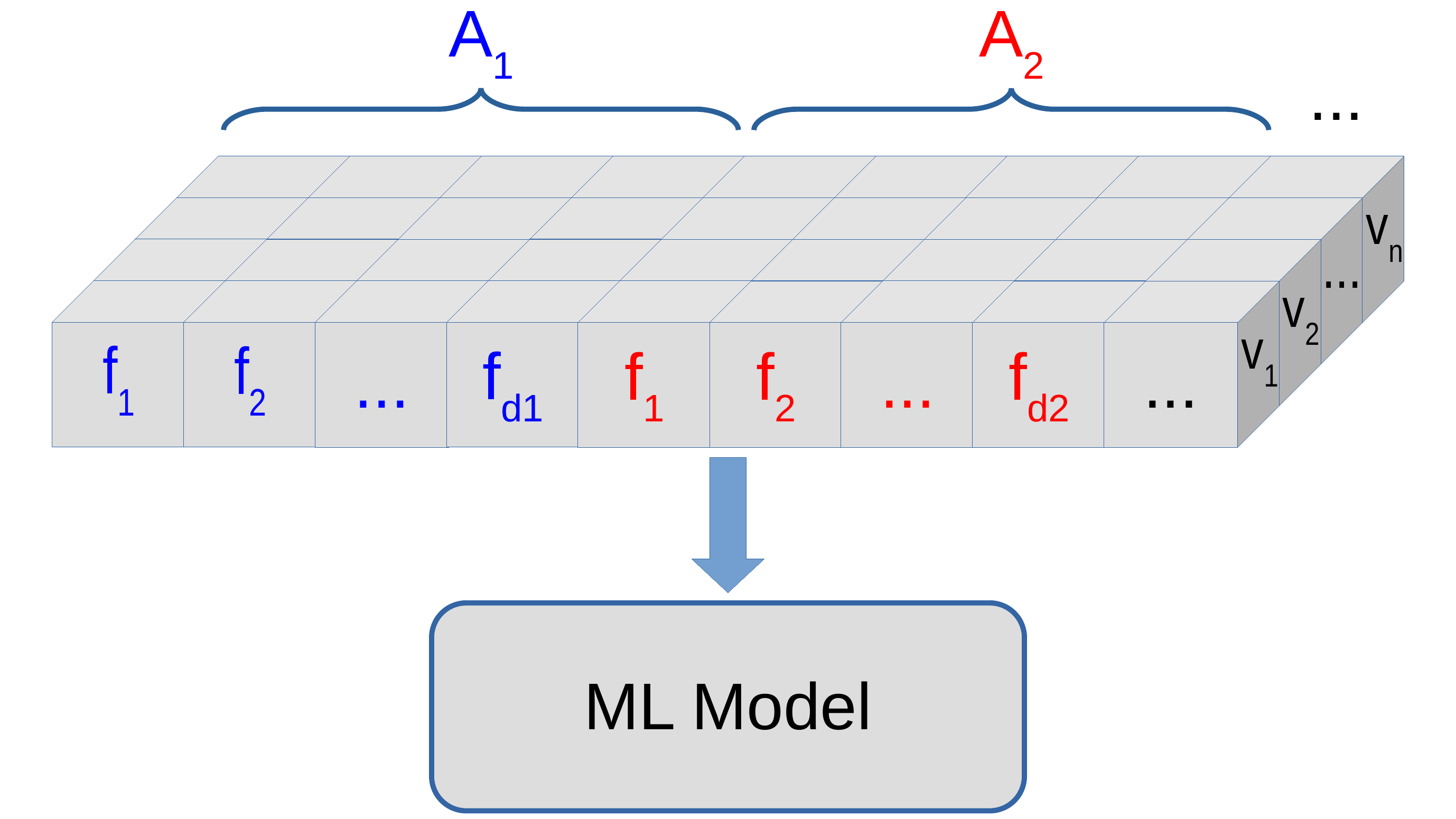}
        \caption{Model design of HoloClean.}
        \label{subfig:hc}
    \end{subfigure}
    \begin{subfigure}[c]{0.50\linewidth}
        \centering
        \includegraphics[height=4cm]{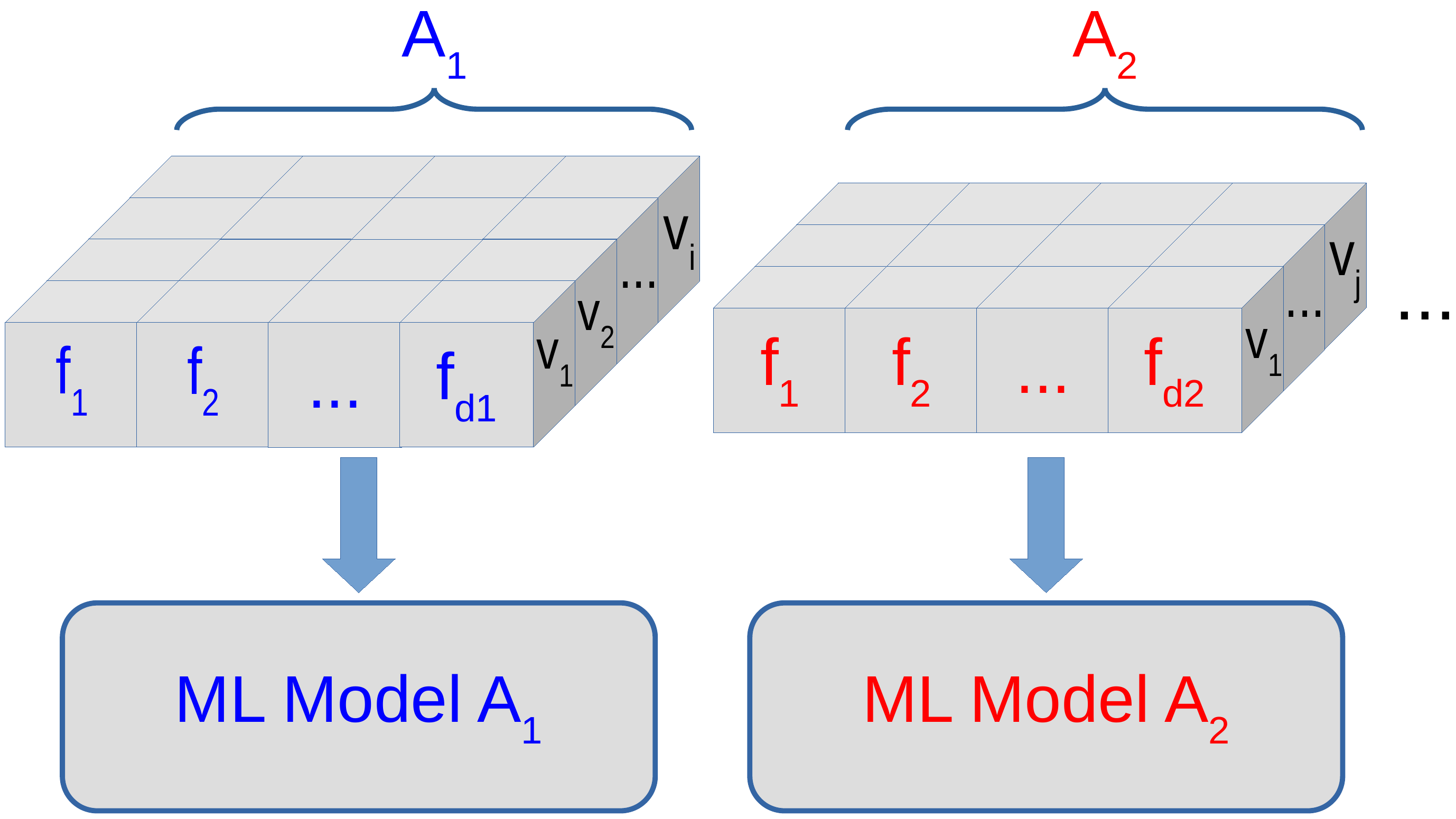}
        \caption{Proposed model design.}
        \label{subfig:attribute-based}
    \end{subfigure}
    \caption{Comparison of ML model designs.}
    \label{fig:models}
\end{figure}

We extended the model design of HoloClean to employ multiple models, one per attribute, as presented in Figure \ref{fig:models}.
One of the benefits enabled by this strategy is the feature vector size reduction.
The first dimension of the tensors is reduced because the maximum number of domain values now considers cells from a single attribute.
Figure \ref{subfig:hc} shows that the single-model design requires reserving more positions in the tensors, in order to accommodate the largest domain size across the whole dataset, $v_{n}$.
Figure \ref{subfig:attribute-based} shows that employing multiple models, on the other hand, allows reserving less positions in the tensors, corresponding to the largest domain size of each attribute, namely $v_{i}$ for attribute $A_{1}$ and $v_{j}$ for attribute $A_{2}$.
In both designs, a mask is applied to disregard unused positions in the tensors for cells having smaller domain sizes, but nevertheless those positions are kept in memory.
Reserving less positions in the first dimension is fundamental to reduce memory consumption.
Our strategy also decreases the second dimension by storing a single feature set (for the corresponding attribute), rather than one feature set per attribute like in the previous model design, thus saving even more memory.
In addition to the tensor size reduction, our model design brings the benefit of a fine-grained model maintenance.
That is, it allows monitoring data variations in each attribute independently, so retraining can occur only when needed for specific models.

\subsection{Training Skipping Strategy}

We propose a strategy for monitoring joint attribute distributions, in order to identify significant changes in data distributions that might require model retraining.
This strategy sharply reduces computational costs, as the decision of whether to retrain the model relies on the available space of attribute values, before starting the feature generation step.

Given a batch $D_{m}$, our strategy computes the joint distribution of each attribute pair $A_{i}, A_{j}$ in the relation, $i \neq j$, considering their co-occurrence frequencies across batches $D_{1..m}$.
Such a set of joint distributions is a simplification of the joint distribution over all attributes, and it is more compact and faster to compute.
At the first batch $D_{1}$, our strategy trains all models.
For the remaining batches, our strategy works as follows.
It starts by fetching the training sets and joint distributions, for each attribute, from the last batch at which the model for $A_{i}$ was trained.
That is, considering that the model was trained at batch $D_h, 1 \leq h < m$, the training set and the joint distributions refer to batches $D_{1..h}$.
Then, the decision of whether to train the model for attribute $A_{i}$ is based on the KL divergence from each joint distribution (with the other attributes $A_{j}$, $i \neq j$) at the last batch that training occurred (over $D_{1..h}$), which we denote $P_{h}(A_{i}, A_{j})$, to the corresponding joint distribution at the current batch (over $D_{1..m}$), which we denote $P_{m}(A_{i}, A_{j})$.
If a KL divergence $KL(P_{m}, P_{h})$ is greater than a user-defined threshold $\epsilon_{kl}$, the model should be retrained.
The intuition is that a significant variation in the distributions was identified, and thus the data within batches $D_{(h+1)..m}$ contain new information that should be taken into account to improve the model.
Whenever a model is (re)trained, the corresponding joint distributions are saved for future reference.

We propose two variants for this strategy.
The first one, denoted by \emph{individual KL} (iKL), triggers model retraining for attribute $A_{i}$ if it detects that any KL divergence value is greater than the threshold provided, i.e. $KL(P_{m}(A_{i}, A_{j}), P_{h}(A_{i}, A_{j})) > \epsilon_{kl}, i \neq j$.
This variant is quite sensitive to distribution changes, as a single KL divergence greater than the given threshold, among all attributes $A_{j}$ other than $A_{i}$, is considered to potentially impact repair accuracy.
The second variant, called \emph{weighted KL} (wKL), considers an aggregate value instead of the individual KL divergence.
Specifically, each individual KL divergence is weighted by the correlation strength of the corresponding attribute pair, and the resulting aggregate value is compared to the user-defined threshold.
In this variant, the model corresponding to attribute $A_{i}$ is retrained if $wKL(A_{i}) > \epsilon_{kl}$, and $wKL(A_{i})$ is computed via the following equation:

\begin{equation}
wKL(A_i) = \frac{1}{N-1} \sum_{j \neq i} [ KL(P_{m}(A_{i}, A_{j}), P_{h}(A_{i}, A_{j}) \cdot corr(A_{i}, A_{j}) ],
\end{equation}

\noindent
where $N$ is the number of attributes in relation $D$, used as the normalization factor, and $corr(A_{i}, A_{j})$ is the correlation strength of attributes $A_{i}, A_{j}$.

As presented in Section \ref{sec:experiments}, both variants of our strategy enable significant speed-ups in execution time, with a reduction in repair accuracy that can be tuned by the user-defined threshold $\epsilon_{kl}$.

\section{Experiments}
\label{sec:experiments}

We evaluated four incremental data cleaning approaches: two competitors, based on the original HoloClean system, and two variants of our proposal, all of them employing the same fundamental components (i.e. error detectors, featurizers, and ML model type).

\subsection*{Evaluated Approaches}

\begin{figure}[!ht]
    \centering
    \includegraphics[width=0.70\textwidth]{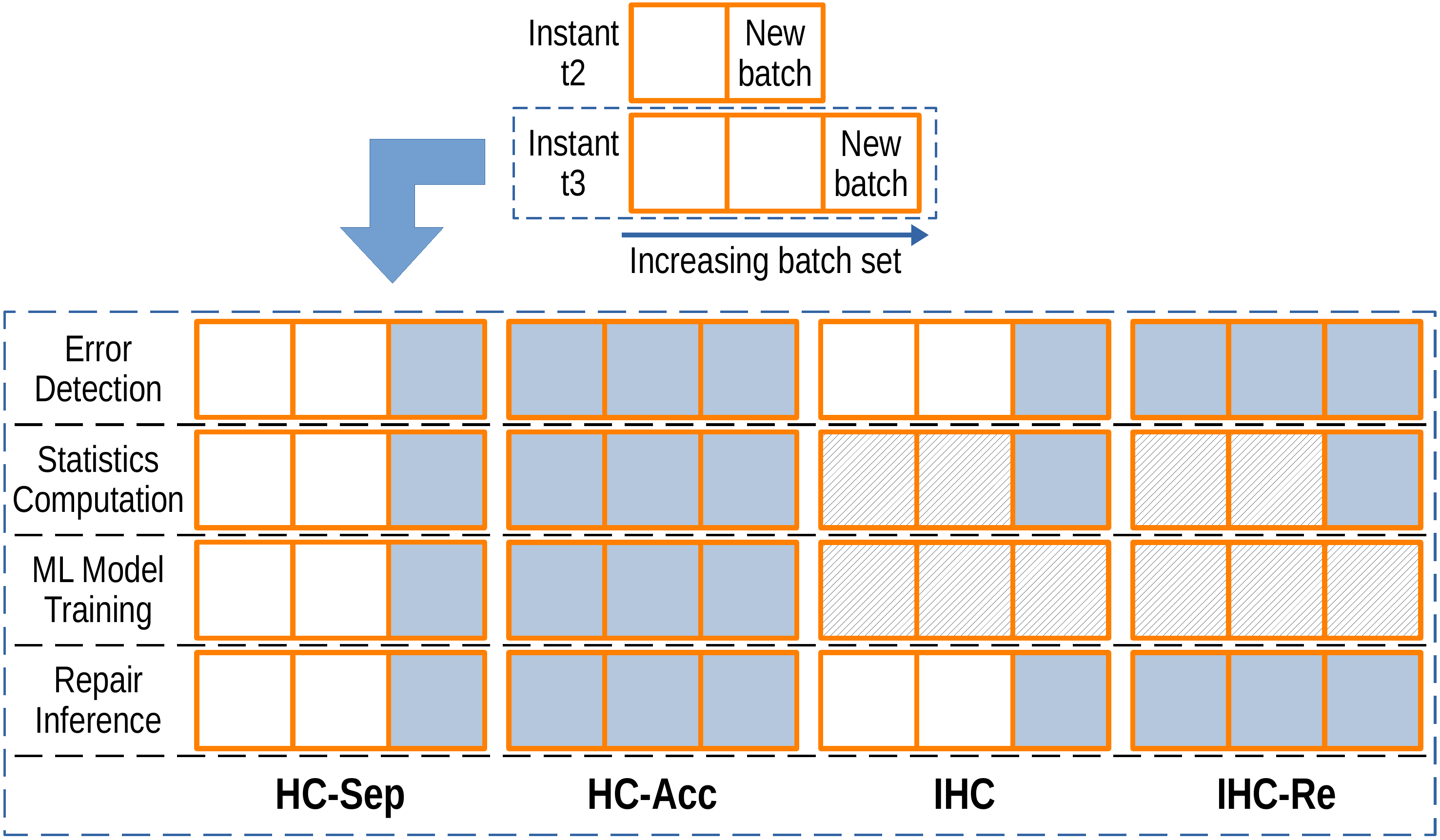}
    \caption{Overview of the evaluated approaches. On the lower section of the figure, the batches at instant $t3$ are shown as squares with the following notation: \textbf{(i)} blank squares are untouched batches; \textbf{(ii)} filled squares are processed batches; and \textbf{(iii)} hatched squares are batches that are more quickly processed/inspected.}
    \label{fig:evaluated-approaches}
\end{figure}

The approaches are shown in Figure \ref{fig:evaluated-approaches} in terms of how they handle an incoming batch when performing error detection, statistics computation, model training, and repair inference.

\textbf{\hca.}
This approach executes HoloClean for each data batch separately.
That is, each batch is handled as a brand new dataset, without the notion of a single dataset evolving.
Therefore, feature generation and model training are performed from scratch for every batch, disregarding the knowledge obtained in previous batches.
There is no information maintenance between batches, and this approach processes the data from a single batch at a time.
Thus, this approach tends to be the fastest among the evaluated ones and to incur little memory consumption, despite employing a single ML model for all attributes.
Conversely, it tends to have the lowest repair quality, as it lacks an increasingly global view of the data.

\textbf{\hcb.}
This approach employs the entire process of HoloClean as well, but it accumulates incoming batches as an evolving dataset.
Hence, instead of processing the data batches in isolation, \hcb appends each new data batch to all the data seen so far and processes the entire dataset from scratch.
Consequently, after appending the last incoming batch, this approach ends up processing the full dataset.
Therefore, among the evaluated approaches, \hcb is the most time- and memory-consuming one, hence being potentially unsuitable in a hypothetical scenario with limited computational resources.
Conversely, it tends to have the highest repair quality as it increasingly gets a global view of the dataset, being able to generate comprehensive data features, as well as to benefit from previous repairs while adjusting features and the model.

\textbf{\ihc.}
This approach is our basic proposal.
It consists of an incremental version of HoloClean that cleans the incoming ($m$-th) batch with ML models trained using all batches seen so far, based on evolving features generated from statistics that are incrementally updated at each new data batch (hence the hatched squares in Figure \ref{fig:evaluated-approaches} for ``Statistics Computation'').
This approach was evaluated for both variants of the proposed training skipping strategy (which explains the hatched squares in Figure \ref{fig:evaluated-approaches} for ``ML Model Training''), referred to as IHC-iKL$\epsilon_{kl}$ and IHC-wKL$\epsilon_{kl}$ accordingly, where $\epsilon_{kl} \in \{0.1, 0.05, 0.01\}$.

\textbf{\ihcre.}
This approach is a variant of our basic proposal.
Its main difference from \ihc is that, when it receives the $m$-th batch, it seeks to repair errors spotted across batches $1..m$, i.e. fixing errors spotted in cells from previous batches plus errors from the current batch.
An intuition behind \ihcre is that certain repairs previously made in earlier batches might not have been successful.
Furthermore, certain errors might not have been detected in previous batches because the available tuples at that moment were not enough to identify certain violations.
Therefore, unlike \ihc, this variant does not assume that the previous repairs fully cleaned the data.
Rather, it assumes that previous repairs turned the data into the cleanest state possible, given the available features and training data.
Overall, \ihcre presents higher repair quality than \ihc does, but requires longer execution times and a greater memory consumption.
Just like \ihc, this approach also updates statistics incrementally and applies our training skipping strategy, thus performing feature generation and model training more quickly than \hcb (and hence the hatched squares in Figure \ref{fig:evaluated-approaches} for \ihcre as well).

We employed three error detectors:
\textbf{(i)} Null Detector;
\textbf{(ii)} Constraint Violation Detector, which looks for Denial Constraints, hence being referred to as DC Detector in the experimental evaluation; and
\textbf{(iii)} Perfect Detector, which emulates the ideal situation where the set of dirty cells detected is exactly the set of true errors in the dataset, obtained from the ground-truth data.
Employing the Perfect Detector allows decoupling data repair from error detection, and this is useful for evaluating the approaches solely on data repair, which is the main focus of this work.

The implementation was based on PyTorch and PostgreSQL, and the experiments were carried out in a machine equipped with a 12-core Intel Xeon E5-2603 1.60GHz, 64GB 2133MHz RAM memory, 7200RPM SATA 6Gbps HDD, and GNU/Linux Ubuntu 14.04.6 LTS operating system.

\subsection*{Datasets}

We employed three datasets for the reported experiments, which are described in Table \ref{tab:datasets}.
We show the cardinality and the dimensionality of each dataset, followed by the error rate and the error detector(s) employed.

\begin{table}[htb]
    \centering
    \caption{Datasets used in the experiments.}
    \begin{tabular}{lcccc}
        \toprule
        \multirow{2}{*}{Dataset} & \# of  & \# of & Error       & Error       \\
                                 & tuples & attrs & rate        & detector(s) \\
        \midrule
        Hospital                 & 1,000  & 17    & $\sim$3\%   & Null + DC   \\
        Food                     & 5,000  & 13    & $\sim$0.5\% & Null + DC   \\
        Soccer-12.5k01pct        & 12,500 & 10    & 0.1\%       & Perfect     \\
        Soccer-12.5k1pct         & 12,500 & 10    & 1\%         & Perfect     \\
        Soccer-12.5k10pct        & 12,500 & 10    & 10\%        & Perfect     \\
        Soccer-25k01pct          & 25,000 & 10    & 0.1\%       & Perfect     \\
        Soccer-25k1pct           & 25,000 & 10    & 1\%         & Perfect     \\
        Soccer-25k10pct          & 25,000 & 10    & 10\%        & Perfect     \\
        Soccer-50k01pct          & 50,000 & 10    & 0.1\%       & Perfect     \\
        Soccer-50k1pct           & 50,000 & 10    & 1\%         & Perfect     \\
        Soccer-50k10pct          & 50,000 & 10    & 10\%        & Perfect     \\
        \bottomrule
    \end{tabular}
    \label{tab:datasets}
\end{table}

The Hospital dataset is widely employed in the data cleaning literature and has hospital census information.
The errors are artificially inserted typos.
Food has data about eating establishments in Chicago.
The errors are inconsistencies of values between pairs of tuples, and the ground truth was manually generated by fixing a small subset of the existing errors.
Soccer is a synthetic dataset about soccer players and their teams, with errors injected by the tool BART \cite{Arocena2015BART}.
We used a subset of the example available on BART's website\footnote{http://db.unibas.it/projects/bart/}.
For all experiments, the datasets were loaded in 100 sequential batches of 1\% the dataset size.

We present the DCs for the Hospital and Food datasets, where $D$ refers to the corresponding relation of the dataset.
The DCs for the Hospital dataset are the following.
\allowdisplaybreaks
\begin{flalign*}
    &dc_{1}: \forall ( t_{\alpha}, t_{\beta} ) \in D, ~ \neg(
        \begin{aligned}[t]
            &t_{\alpha}.condition = t_{\beta}.condition ~ \wedge &\\
            &t_{\alpha}.measure\_name = t_{\beta}.measure\_name ~ \wedge &\\
            &t_{\alpha}.hospital\_type \neq t_{\beta}.hospital\_type)
        \end{aligned} &\\
    &dc_{2}: \forall ( t_{\alpha}, t_{\beta} ) \in D, ~ \neg(
        \begin{aligned}[t]
            &t_{\alpha}.hospital\_name = t_{\beta}.hospital\_name ~ \wedge ~ t_{\alpha}.zip\_code \neq t_{\beta}.zip\_code)
        \end{aligned} &\\
    &dc_{3}: \forall ( t_{\alpha}, t_{\beta} ) \in D, ~ \neg(
        \begin{aligned}[t]
            &t_{\alpha}.hospital\_name = t_{\beta}.hospital\_name ~ \wedge &\\
            &t_{\alpha}.phone\_number \neq t_{\beta}.phone\_number)
        \end{aligned} &\\
    &dc_{4}: \forall ( t_{\alpha}, t_{\beta} ) \in D, ~ \neg(
        \begin{aligned}[t]
            &t_{\alpha}.measure\_code = t_{\beta}.measure\_code ~ \wedge &\\
            &t_{\alpha}.measure\_name \neq t_{\beta}.measure\_name)
        \end{aligned} &\\
    &dc_{5}: \forall ( t_{\alpha}, t_{\beta} ) \in D, ~ \neg(
    \begin{aligned}[t]
        &t_{\alpha}.measure\_code = t_{\beta}.measure\_code ~ \wedge ~ t_{\alpha}.state\_avg \neq t_{\beta}.state\_avg)
    \end{aligned} &\\
    &dc_{6}: \forall ( t_{\alpha}, t_{\beta} ) \in D, ~ \neg(
    \begin{aligned}[t]
        &t_{\alpha}.provider = t_{\beta}.provider ~ \wedge ~ t_{\alpha}.hospital\_name \neq t_{\beta}.hospital\_name)
    \end{aligned} &\\
    &dc_{7}: \forall ( t_{\alpha}, t_{\beta} ) \in D, ~ \neg(
    \begin{aligned}[t]
        &t_{\alpha}.measure\_code = t_{\beta}.measure\_code ~ \wedge ~ t_{\alpha}.condition \neq t_{\beta}.condition)
    \end{aligned} &\\
    &dc_{8}: \forall ( t_{\alpha}, t_{\beta} ) \in D, ~ \neg(
    \begin{aligned}[t]
        &t_{\alpha}.hospital\_name = t_{\beta}.hospital\_name ~ \wedge ~ t_{\alpha}.address1 \neq t_{\beta}.address1)
    \end{aligned} &\\
    &dc_{9}: \forall ( t_{\alpha}, t_{\beta} ) \in D, ~ \neg(
    \begin{aligned}[t]
        &t_{\alpha}.hospital\_name = t_{\beta}.hospital\_name ~ \wedge &\\
        &t_{\alpha}.hospital\_owner \neq t_{\beta}.hospital\_owner)
    \end{aligned} &\\
    &dc_{10}: \forall ( t_{\alpha}, t_{\beta} ) \in D, ~ \neg(
    \begin{aligned}[t]
        &t_{\alpha}.hospital\_name = t_{\beta}.hospital\_name ~ \wedge ~ t_{\alpha}.provider \neq t_{\beta}.provider)
    \end{aligned} &\\
    &dc_{11}: \forall ( t_{\alpha}, t_{\beta} ) \in D, ~ \neg(
    \begin{aligned}[t]
        &t_{\alpha}.hospital\_name = t_{\beta}.hospital\_name ~ \wedge &\\
        &t_{\alpha}.phone\_number = t_{\beta}.phone\_number ~ \wedge &\\
        &t_{\alpha}.hospital\_owner = t_{\beta}.hospital\_owner ~ \wedge &\\
        &t_{\alpha}.state \neq t_{\beta}.state)
    \end{aligned} &\\
    &dc_{12}: \forall ( t_{\alpha}, t_{\beta} ) \in D, ~ \neg(
    \begin{aligned}[t]
        &t_{\alpha}.city = t_{\beta}.city ~ \wedge ~ t_{\alpha}.county\_name \neq t_{\beta}.county\_name)
    \end{aligned} &\\
    &dc_{13}: \forall ( t_{\alpha}, t_{\beta} ) \in D, ~ \neg(
    \begin{aligned}[t]
        &t_{\alpha}.zip\_code = t_{\beta}.zip\_code ~ \wedge ~ t_{\alpha}.emergency \neq t_{\beta}.emergency)
    \end{aligned} &\\
    &dc_{14}: \forall ( t_{\alpha}, t_{\beta} ) \in D, ~ \neg(
    \begin{aligned}[t]
        &t_{\alpha}.hospital\_name = t_{\beta}.hospital\_name ~ \wedge ~ t_{\alpha}.city \neq t_{\beta}.city)
    \end{aligned} &\\
    &dc_{15}: \forall ( t_{\alpha}, t_{\beta} ) \in D, ~ \neg(
    \begin{aligned}[t]
        &t_{\alpha}.measure\_name = t_{\beta}.measure\_name ~ \wedge &\\
        &t_{\alpha}.measure\_code \neq t_{\beta}.measure\_code)
    \end{aligned} &\\
\end{flalign*}

\noindent
The DCs for the Food dataset are the following.
\allowdisplaybreaks
\begin{flalign*}
    &dc_{16}: \forall ( t_{\alpha}, t_{\beta} ) \in D, ~ \neg(
        \begin{aligned}[t]
            &t_{\alpha}.dba\_name = t_{\beta}.dba\_name ~ \wedge &\\
            &t_{\alpha}.address = t_{\beta}.address ~ \wedge &\\
            &t_{\alpha}.facility\_type \neq t_{\beta}.facility\_type)
        \end{aligned} &\\
    &dc_{17}: \forall ( t_{\alpha}, t_{\beta} ) \in D, ~ \neg(
        \begin{aligned}[t]
            &t_{\alpha}.zip\_code = t_{\beta}.zip\_code ~ \wedge ~ t_{\alpha}.city \neq t_{\beta}.city)
        \end{aligned} &\\
    &dc_{18}: \forall t_{\alpha} \in D, \neg(t_{\alpha}.facility\_type = \text{`empty'}) &\\
    &dc_{19}: \forall t_{\alpha} \in D, \neg(t_{\alpha}.city = \text{`empty'}) &\\
    &dc_{20}: \forall t_{\alpha} \in D, \neg(t_{\alpha}.zip\_code = \text{`empty'}) &
\end{flalign*}

\subsection{Performance Overview}

\begin{figure}[!ht]
    \centering
    \includegraphics[width=0.90\textwidth]{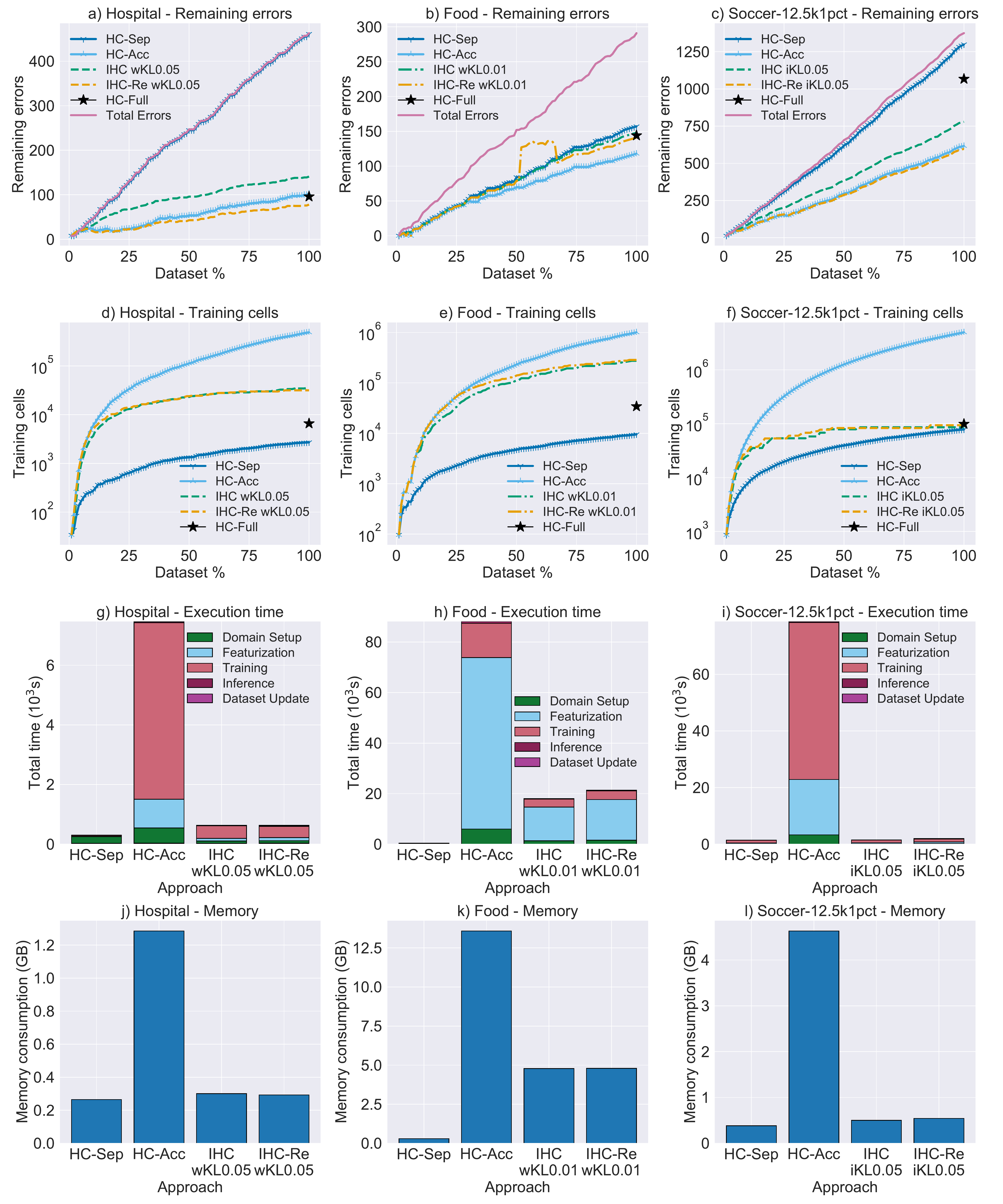}
    \caption{Overview of repair quality and computational efficiency of the proposed approaches \ihc and \ihcre against competitors \hca and \hcb.}
    \label{fig:overall-performance}
\end{figure}

This section shows the experiments performed to evaluate the four approaches on repair quality and computational efficiency.
The results are shown in Figure \ref{fig:overall-performance} and discussed as follows.

With respect to \textbf{data cleaning effectiveness}, the charts at the first row show the number of accumulated errors at each incoming batch, as well as the total number of remaining errors after executing each approach.
For reference, the figure depicts a star marker representing the number of non-repaired errors after executing HoloClean over the whole dataset (\hcf), although this approach does not apply to our scenario of interest where datasets are loaded in batches.
Figure \ref{fig:overall-performance}a shows that the approach \hca barely repaired the dataset, whereas the other approaches achieved cleaning rates greater than 70\%.
This behavior was expected, as the batch size is rather small for this dataset, and it also shows that updating features incrementally is important for the whole process.
Compared to \hcf, the \hcb approach was able to fix a similar number of errors.
Our approach \ihcrew05 was superior, as it leverages the repairs performed in previous batches (which affect the features and models at posterior batches), in addition to employing attribute-based models.
With respect to Food (Figure \ref{fig:overall-performance}b), \hca was quite effective at repairing, being slightly surpassed by \ihcw01 and \ihcrei01.
\ihcrew01 enabled a superior result compared to \hcf, although having presented a prediction problem between batches 51 and 67, but which was fixed afterwards.
Regarding the Soccer12k01pct dataset (Figure \ref{fig:overall-performance}c), \hca performed fewer repairs once again.
Interestingly, the number of repairs provided by the remaining approaches was much superior compared to the execution over the entire dataset, which shows that, for this dataset, maintaining the models and features incrementally was significantly more effective.
Overall, we can observe that \ihcrei05 was the better approach, followed closely by \hcb and a bit farther away by \ihci05.

Focusing on \textbf{model training effort}, the second row of charts in Figure \ref{fig:overall-performance} shows the accumulated number of instances used for training, which reflects the total effort made by each approach.
Regarding the Hospital dataset (Figure \ref{fig:overall-performance}d), the proposed approaches (\ihcw05 and \ihcrew05) used one order of magnitude more instances than \hca, as well as one order of magnitude less instances than \hcb (note the logarithmic scale on these charts).
With respect to Food (Figure \ref{fig:overall-performance}e), \ihcw01 and \ihcrew01 got closer to \hcb.
Considering Soccer12.5k1pct, the proposed approaches used a similar number of instances compared to \hca, which shows that our ML models converged more quickly.
For all datasets, we can observe that \hca used less instances than \hcf due to differences both in the number of reported errors and in the number of labelled cells, generated by the \emph{weak-labelling} step of HoloClean for augmenting the training set.

The third row of charts in Figure \ref{fig:overall-performance} shows the total \textbf{execution time} for processing all data batches.
It is clear that the exhaustive execution performed by \hcb is too time-consuming and tends to be unsuitable for large datasets.
Considering both proposed approaches, \ihc enabled the best results in terms of execution time, taking just 2\%--20\% of the time spent by the \hcb competitor.
A particular behavior regarding Food is the high cost for the feature generation step, which surpassed the execution time for model training.
This occurred due to the attribute domains being too large.
Overall, the \hca approach was much faster than the other approaches, in detriment of repair quality.

Finally, with respect to \textbf{memory consumption}, the last row of charts in Figure \ref{fig:overall-performance} shows that our proposals reduced significantly the amount of memory required.
Specifically, the \ihc and \ihcre approaches required 10\%--35\% the maximum amount of memory used by \hcb.

\subsection{Analysis of Attribute-Based Models}

A key point in our proposal is the adoption of attribute-based models.
Both \ihc and \ihcre approaches had similar (and lower) feature generation and training costs compared to \hcb.
Particularly, \ihcre can be seen as a variant of \hcb that employs attribute-based models.
Hence, our expectation was that attribute-based models would provide a similar repair quality compared to the original single-model design, although requiring less memory.

\begin{figure}[!ht]
    \centering
    \includegraphics[width=0.90\textwidth]{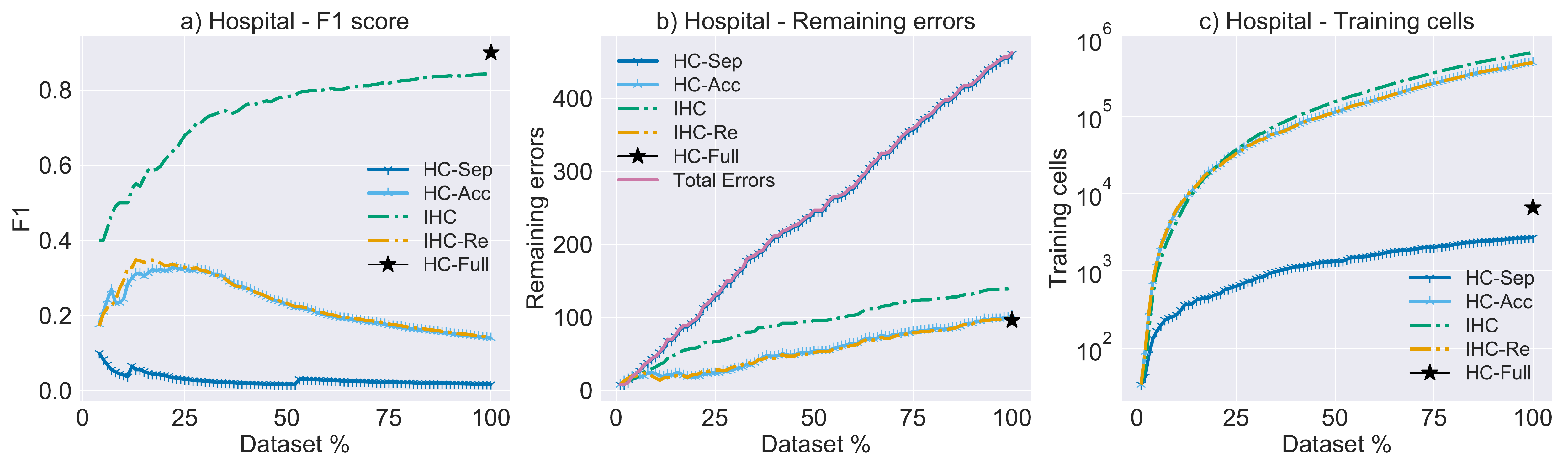}
    \caption{Repair quality and number of training instances using attribute-based vs. single-model approaches.}
    \label{fig:attribute-models-quality}
\end{figure}

Figure \ref{fig:attribute-models-quality}a presents the F1 score of each approach over Hospital.
The figure shows that the repair quality of \ihcre was similar to the one of \hcb, and \ihc had a considerably higher F1 score, close to the one of \hcf.
The attempt to re-repair errors from previous batches decreases the recall in \ihcre and \hcb, as they repeatedly consider cells that could not be repaired previously.
Hence, to effectively compare all the approaches in terms of repair quality, we show the accumulated number of non-repaired errors (Figure \ref{fig:attribute-models-quality}b).
Furthermore, the accumulated number of training instances was similar for approaches \hcb, \ihc, and \ihcre (Figure \ref{fig:attribute-models-quality}c).
Overall, the same behavior occurred for the remaining datasets.

\begin{figure}[!ht]
    \centering
    \includegraphics[width=0.56\textwidth]{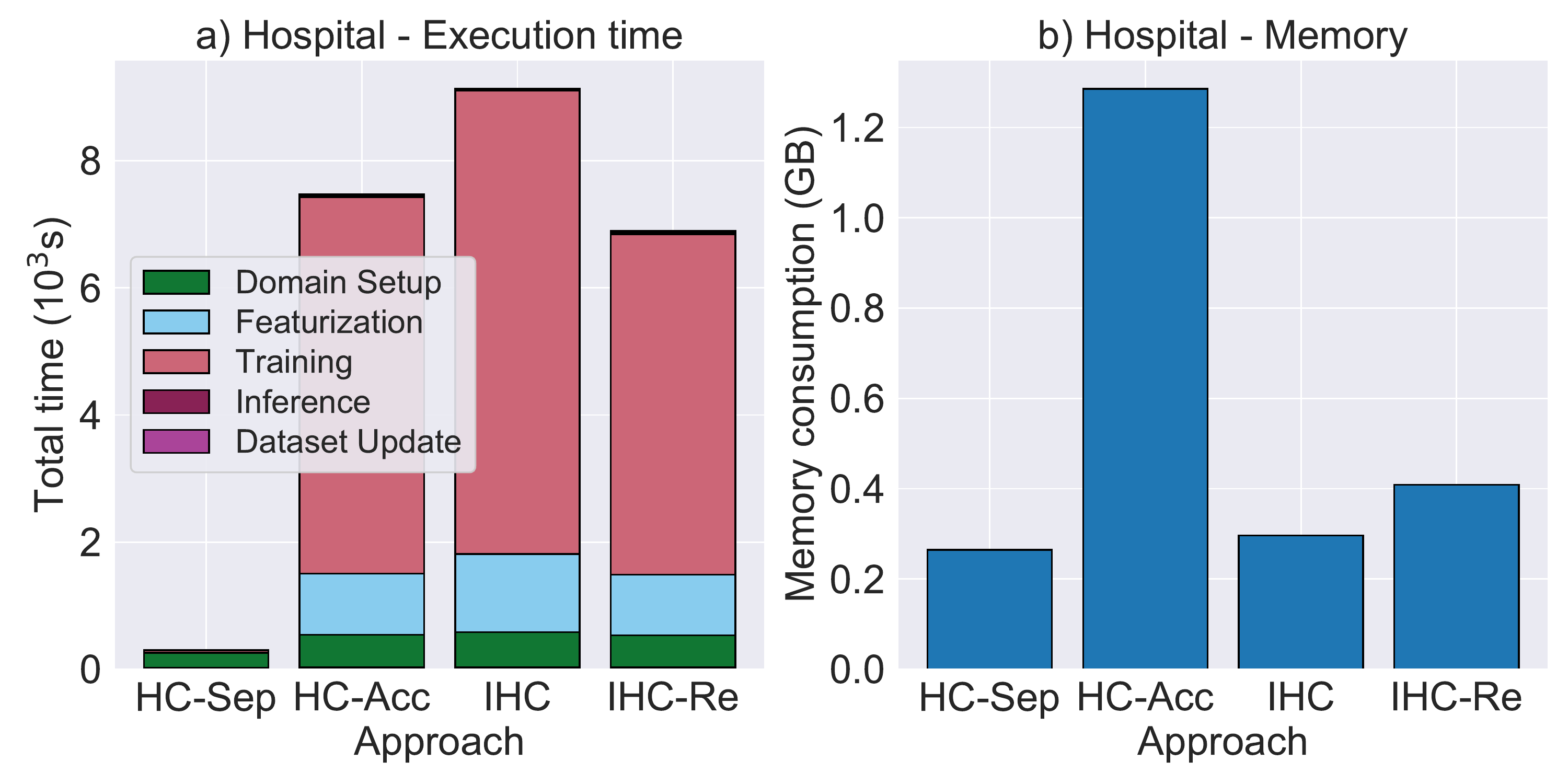}
    \caption{Execution time and memory consumption using attribute-based vs. single-model approaches.}
    \label{fig:attribute-models-time-memory}
\end{figure}

Regarding execution time, Figure \ref{fig:attribute-models-time-memory}a shows that the total times of \ihc and \ihcre were relatively similar to the one of \hcb and quite longer compared to \hca.
For this specific case, the execution time for \ihcre was shorter than for \ihc.
This happened because, while certain dirty cells from initial batches were cleaned when being revisited by \ihcre, \ihc left them untouched.
Therefore, \ihc allowed those cells to conflict with clean cells from the posterior batches, which caused such clean cells to be marked as potential errors.
This fact made \ihc spend more time by needlessly generating features for clean cells and attempting to repair them in posterior batches.
Considering memory consumption, the proposed approaches were way better than \hcb, as shown in Figure \ref{fig:attribute-models-time-memory}b.
Moreover, combining attribute-based models with our strategy for skipping model training enabled a sharp reduction in execution time, as we show in the next section.

\subsection{Analysis of Training Skipping Strategy Variants}

\begin{figure}[!ht]
    \centering
    \includegraphics[width=0.62\textwidth]{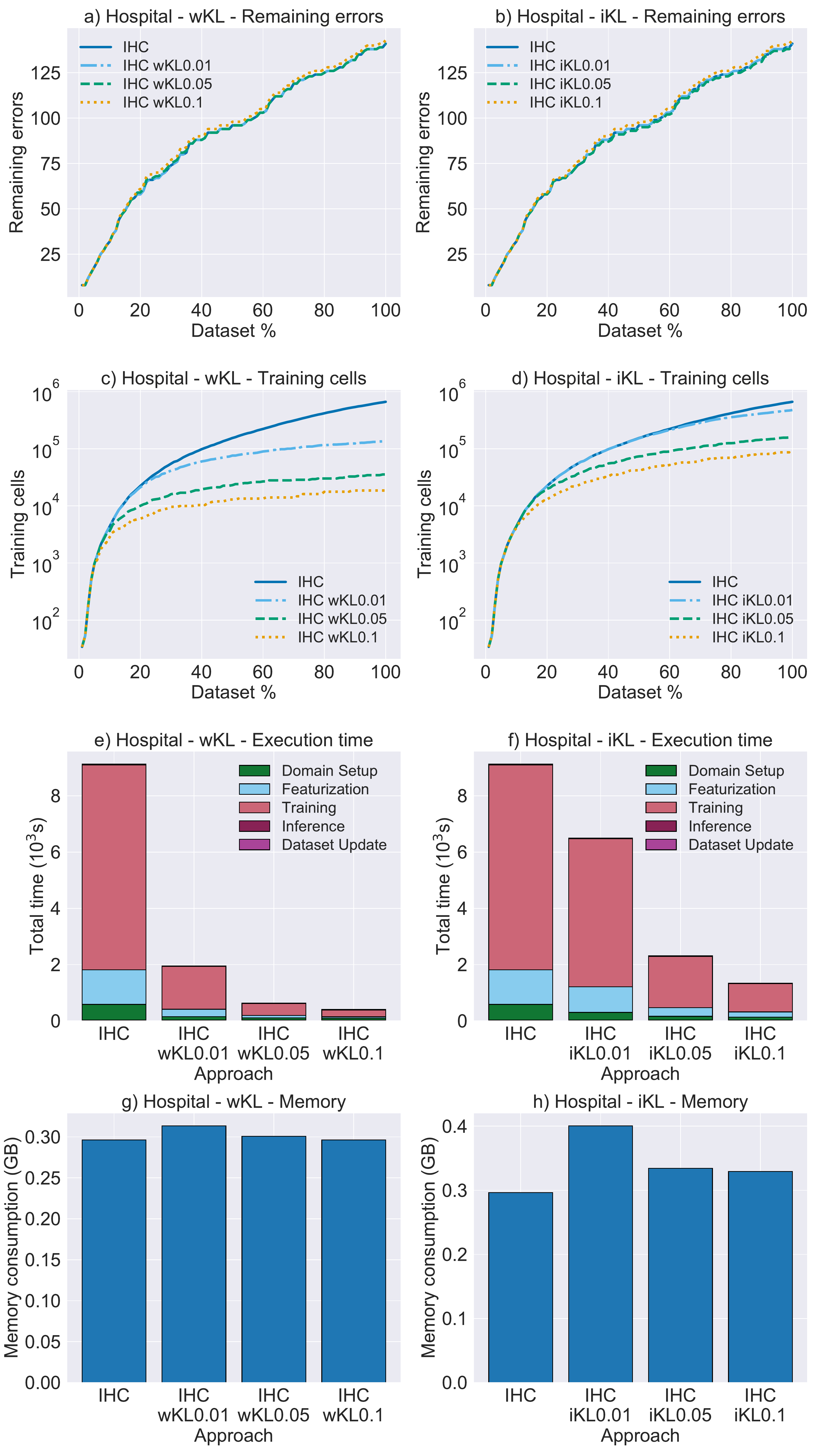}
    \caption{Behavior of threshold variation in training skipping strategy for the Hospital dataset.}
    \label{fig:training-skipping-hospital}
\end{figure}

\begin{figure}[!ht]
    \centering
    \includegraphics[width=0.62\textwidth]{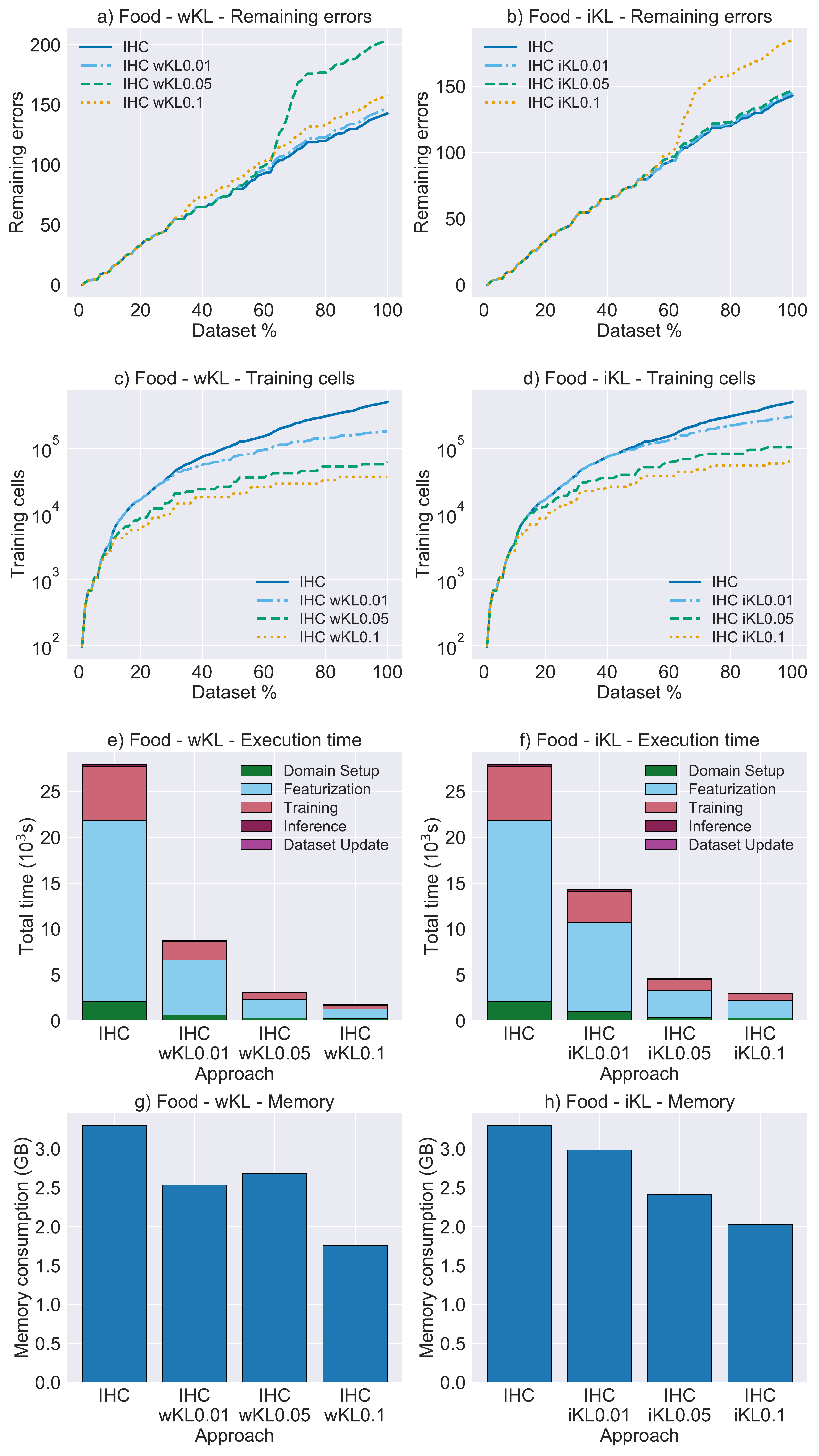}
    \caption{Behavior of threshold variation in training skipping strategy for the Food dataset.}
    \label{fig:training-skipping-food}
\end{figure}

Domain and feature generation are among the most time-consuming tasks in our pipeline.
Therefore, we focused on improving computational performance by attempting to skip these tasks whenever model retraining is not needed.
Indeed, the proposed training skipping strategy led to significant performance improvements.
The strategy reduces execution time according to a threshold parameter, in order to control the resulting impact on repair quality.
This section presents the behavior of our data cleaning approaches when employing the iKL and wKL variants, taking as examples the datasets Hospital (Figure \ref{fig:training-skipping-hospital}) and Food (Figure \ref{fig:training-skipping-food}).

With respect to the Hospital dataset, Figure \ref{fig:training-skipping-hospital} shows that the highest threshold value ($\epsilon_{kl}=0.1$) enabled the fastest execution time, as well as led to just a slight reduction in the number of repairs compared to the other threshold values.
Furthermore, \ihci05 had the highest repair quality, which points out that more training not always leads to more repairs.
We can also observe that the number of training instances and the execution times decrease as the provided threshold increases.
For instance, \ihcw1 spent about 2\% the time taken by \ihc, with a slight difference in the number of repairs.
Regarding memory consumption, there is an overall reduction in the required memory amount as the threshold increases, although such reduction is less significant than the ones in the number of training instances and in execution time.
Overall, we observed that the iKL strategy is more sensitive to changes than wKL, hence leading to model retraining more often.

A similar overall behavior could be seen for the Food dataset (Figure \ref{fig:training-skipping-food}).
However, compared to Hospital, distinct threshold values had a more significant effect on the resulting numbers of repairs, which were significantly more different from each other as the threshold varied.
A particular situation occurred for approaches \ihcw05 and \ihci1, which was a problem between batches 60 and 70 that led to unsatisfactory repairs.
The models regained their performance around batch 70, which could have enabled a delayed repair of the affected cells if the re-repairing mechanism of \ihcre were being employed.

\subsection{Impact of Data Variations on the Evaluated Approaches}

\begin{figure}[!ht]
    \centering
    \includegraphics[width=0.60\textwidth]{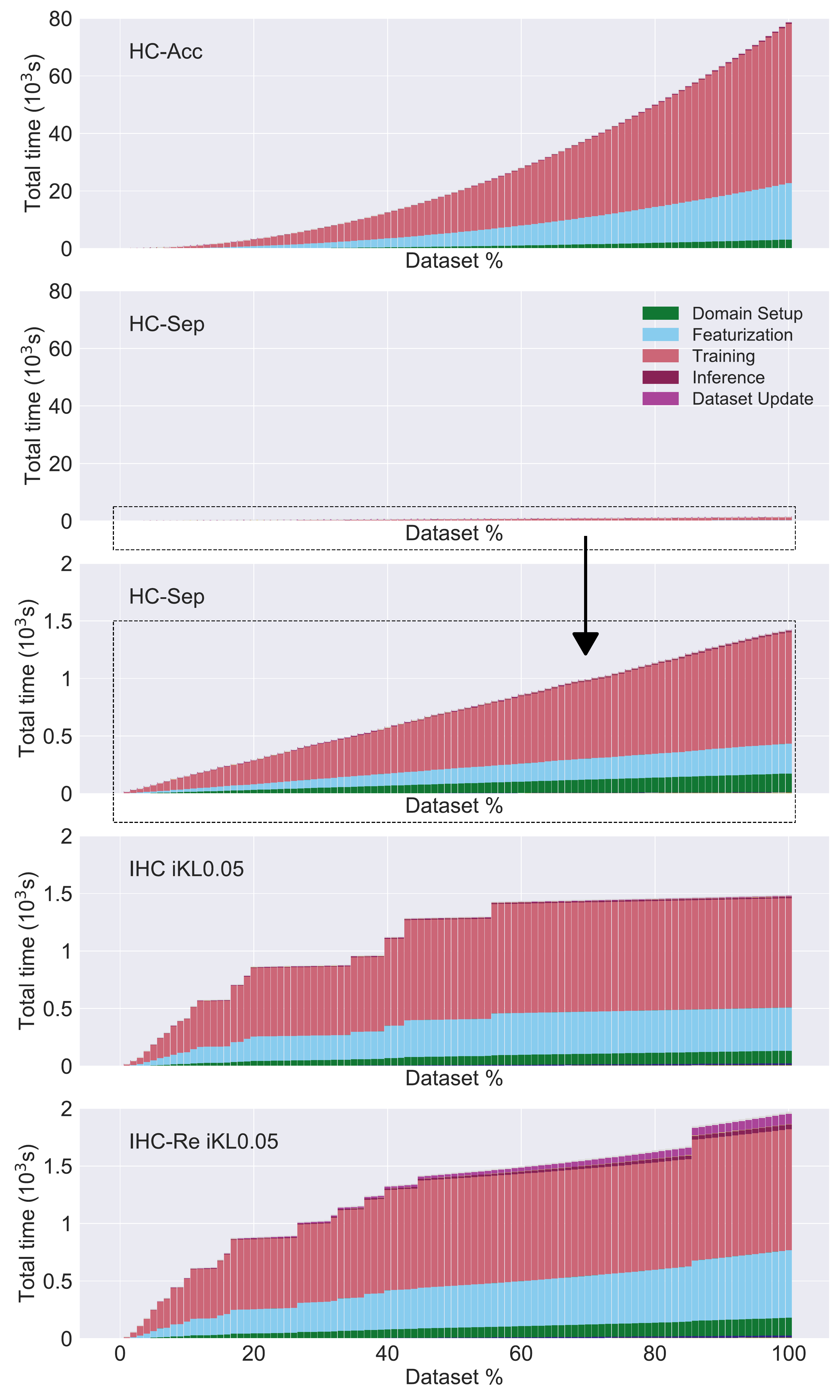}
    \caption{Overview of accumulated execution times for the evaluated approaches over Soccer12.5k1pct.}
    \label{fig:soccer-acc-time-overview}
\end{figure}

This section describes the behavior observed in the proposed approaches, compared to the competitors, as the data vary by size and error rate.
This analysis was based on the Soccer dataset employing the Perfect error detector, in order to isolate the data repairing step and evaluate the parameters in a controlled environment.

We begin the analysis by closely inspecting the accumulated execution times, shown for each evaluated approach in Figure \ref{fig:soccer-acc-time-overview}, as the Soccer12.5k1pct dataset evolves.
Both plots on the upper section of the figure depict the accumulated times for competitors \hcb and \hca respectively.
As expected, the execution time for \hcb increases exponentially as more batches are cleaned, owing to the processing of each batch becoming more comprehensive.
In this experiment, the total execution for \hcb was around $56\times$ longer than for \hca.
Right below in the figure, the accumulated execution times for \hca are presented using a different scale, showing that the growth is linear, since the processing efforts at each new batch are constant for this approach.
With respect to the proposed approaches \ihc and \ihcre employing our training skipping strategy, the execution time evolution is sublinear.
Compared to \hca, we can observe higher amounts of effort in the initial batches.
However, such efforts start to increase more slowly over time, as the ML models become stable and the need for retraining is reduced.
The growth of execution time for \ihcre is greater than for \ihc, due to the costs related to cell domain computation, feature generation, as well as dataset updates on cells from previous batches identified as erroneous.
The (re)training costs in both approaches were almost the same, since the same training skipping strategy and threshold value (iKL0.05) were employed.
Based on this analysis, further results regarding the approach \hcb were not inspected for larger datasets.

\begin{figure}[!ht]
    \centering
    \includegraphics[width=0.90\textwidth]{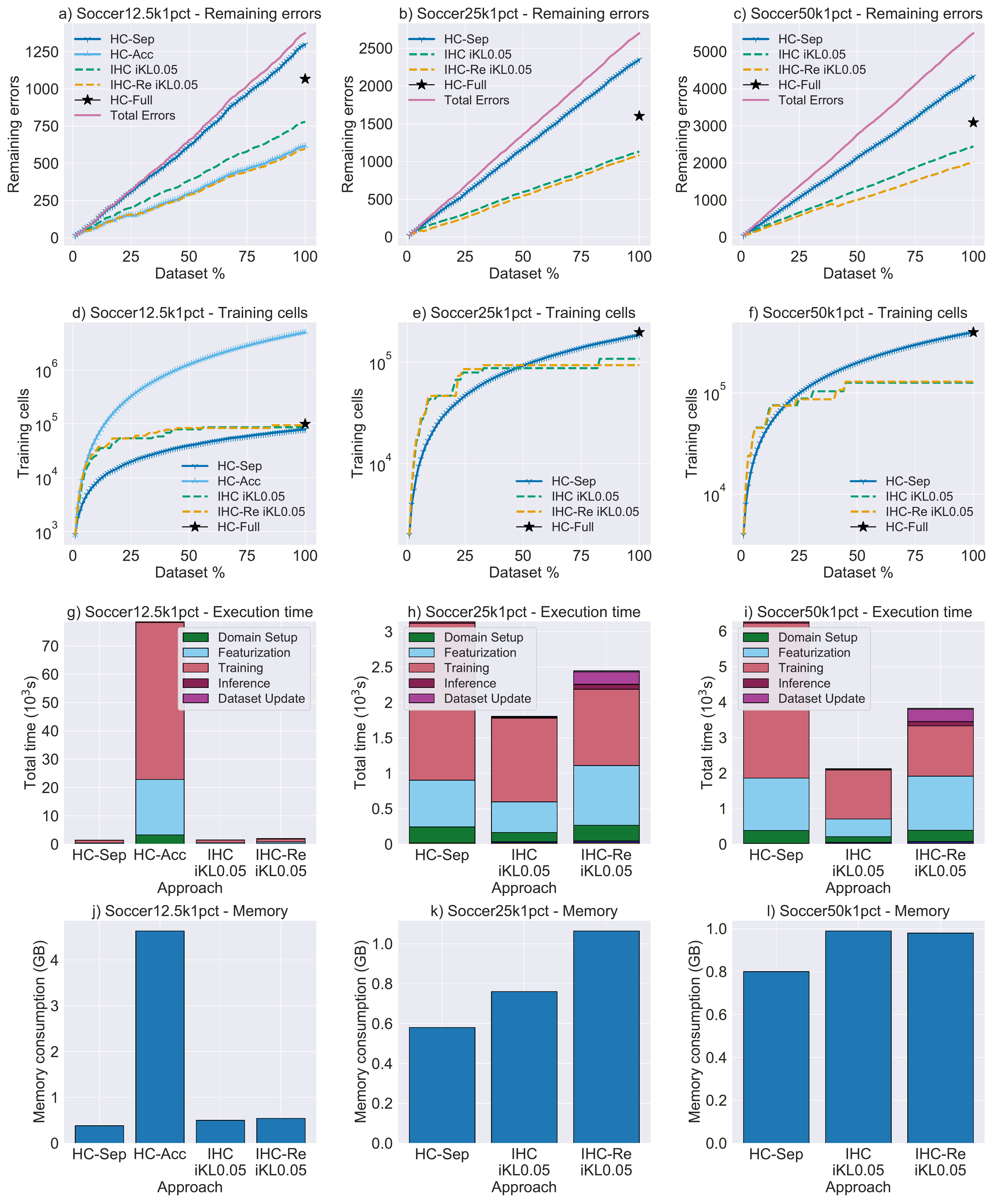}
    \caption{Comparison of approaches with varying dataset sizes and a fixed error rate of 1\%.}
    \label{fig:soccer-varying-size}
\end{figure}

Figure \ref{fig:soccer-varying-size} depicts our analysis on \textbf{varying dataset size}, considering results over the Soccer dataset having different amounts of tuples and a fixed error rate of 1\%.
The results for \hcb are presented just for the dataset with 12,500 tuples, as its costs become prohibitively high over larger datasets.
As we can observe on the first row of plots in the figure, \ihcre is the approach that overall enables the greatest number of repairs, followed by \hcb (when evaluated) and by \ihc.
For all cardinalities, these approaches were also better than executing HoloClean over the full dataset.
The \hca approach presented an increasingly higher repair quality as the dataset size grew larger, since the batches themselves also became larger.
Yet, \hca was still the worst approach.
Considering the number of training instances, represented by the second row in the figure, we can notice that employing the proposed training skipping strategy allowed saving efforts as the dataset grew, since the models were able to stabilize with less processed batches.
With respect to execution time, our approaches increasingly outperformed \hca as the dataset grew larger.
The plots in Figure \ref{fig:soccer-varying-size}h and Figure \ref{fig:soccer-varying-size}i, in turn, depict the overhead of \ihcre in several tasks, due to its re-repairing mechanism.
Finally, the last row of plots show the sublinear memory consumption growth as the dataset evolves.

\begin{figure}[!ht]
    \centering
    \includegraphics[width=0.90\textwidth]{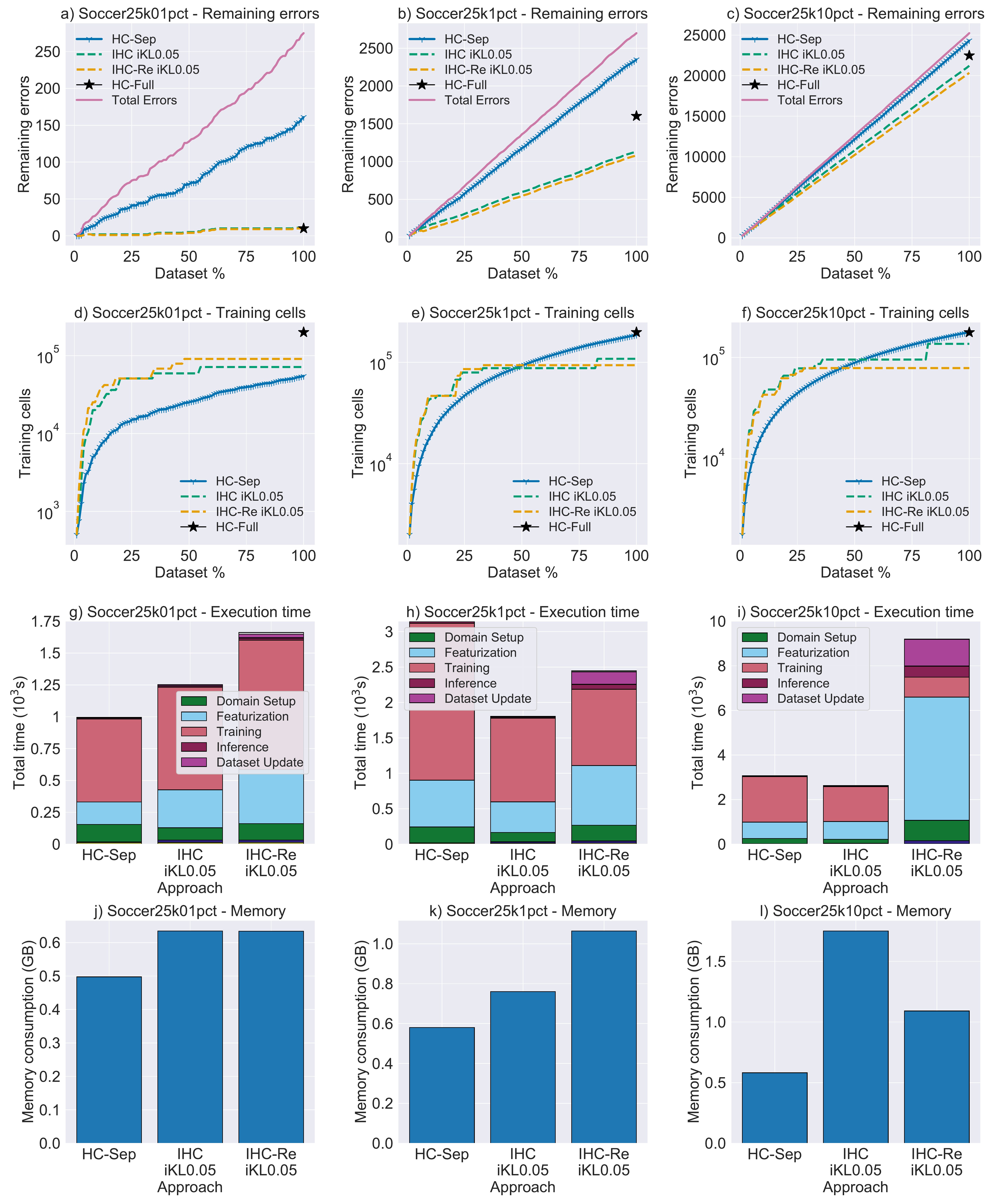}
    \caption{Comparison of approaches over the Soccer25k dataset with varying error rates.}
    \label{fig:soccer-varying-pct}
\end{figure}

Considering \textbf{varying error rate}, Figure \ref{fig:soccer-varying-pct} shows the behavior of each approach over Soccer with 25,000 tuples, using error rates of 0.1\%, 1\%, and 10\%.
In these experiments, the error rate refers to the fraction of \emph{cells} (not \emph{tuples}) that are dirty.
We can observe that all approaches degrade as the error rate increases.
This behavior was expected, as predicting the most likely repair relies on information extracted from the dataset itself and, typically, the greater the noise, the lower the accuracy.
Regarding the number of training instances, our approaches \ihci05 and \ihcrei05 ended up requiring less instances than \hca for the error rates of 1\% and 10\%, which points out that our models stabilized more quickly.
Considering execution time, all three approaches took longer to execute as the error rate increased, and \ihcrei05 was the one that took significantly longer, especially due to featurization.
Overall, we see that memory consumption increased slightly with the error rate.

\begin{figure}[!ht]
    \centering
    \includegraphics[width=0.90\textwidth]{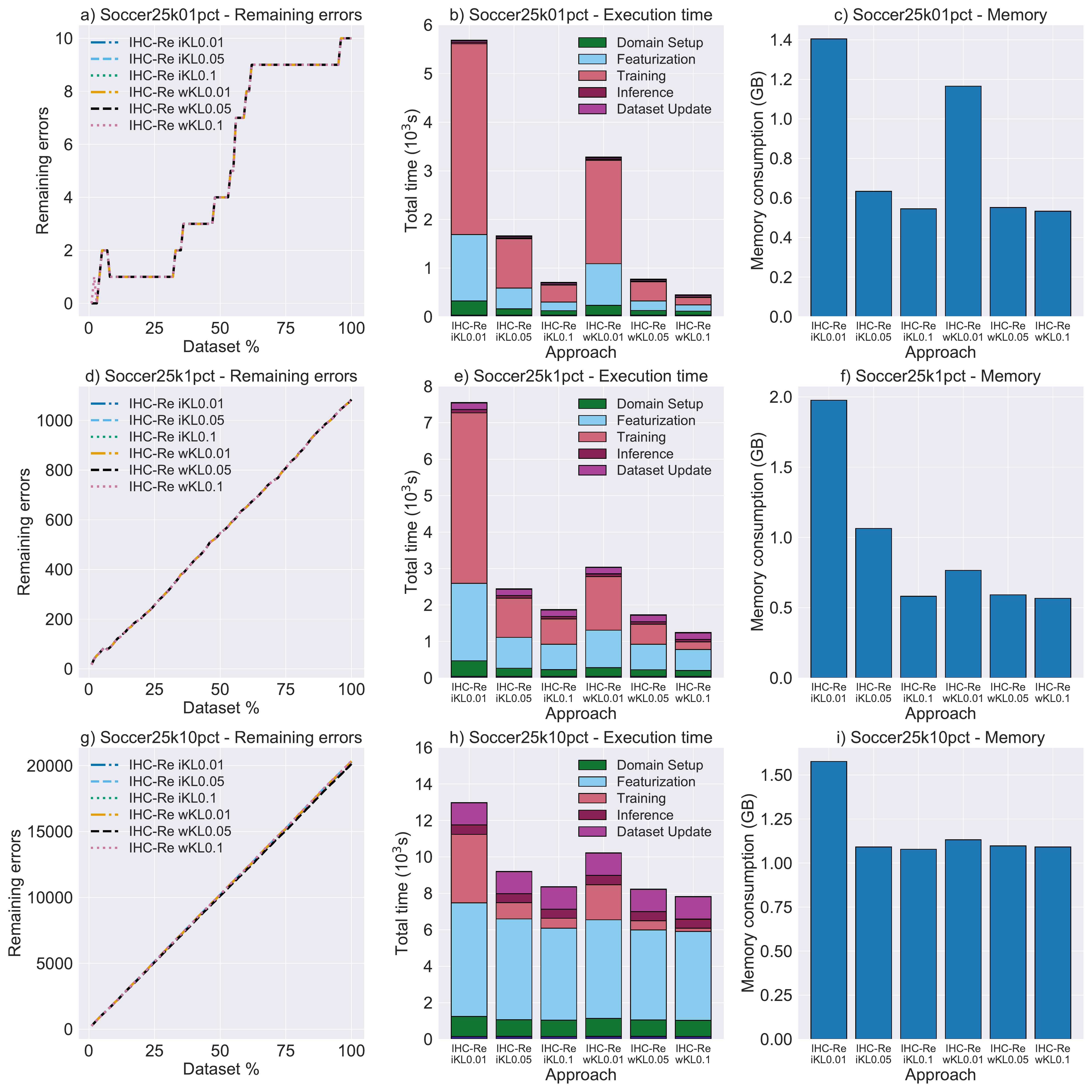}
    \caption{Behavior of training skipping strategy variants in approach \ihcre, considering threshold values 0.01, 0.05, and 0.1, over Soccer25k with varying error rates.}
    \label{fig:ihc-re-skipping-strategy-varying-threshold}
\end{figure}

\begin{figure}[!ht]
    \centering
    \includegraphics[width=\textwidth]{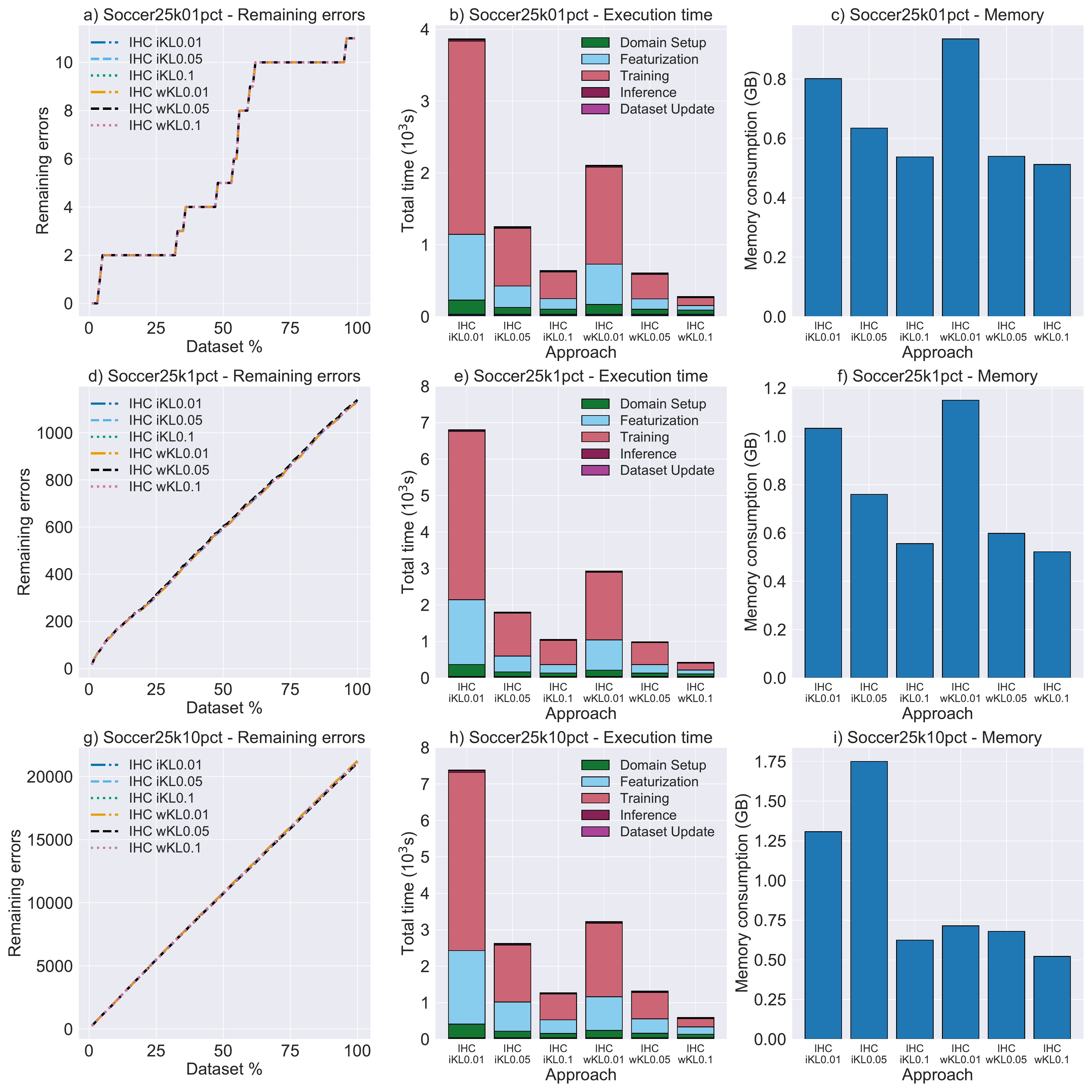}
    \caption{Behavior of training skipping strategy variants in approach \ihc, considering threshold values 0.01, 0.05, and 0.1, over Soccer25k with varying error rates.}
    \label{fig:ihc-skipping-strategy-varying-threshold}
\end{figure}

The last experiment analyzes the approaches with respect to the training skipping strategy variants, which were employed with multiple threshold values over the Soccer25k dataset with distinct error rates.
Figure \ref{fig:ihc-re-skipping-strategy-varying-threshold} and Figure \ref{fig:ihc-skipping-strategy-varying-threshold} show the overall behavior of approaches \ihcre and \ihc, respectively, employing strategy variants iKL and wKL with threshold values 0.01, 0.05, and 0.1.
For all evaluations, the number of repairs was not significantly impacted by the training skipping strategy variant nor by the threshold value.
In contrast, the chosen values were quite relevant when considering total execution time and, at a lowest degree, memory consumption.
The total execution times in \ihcre were incrementally affected by higher error rates (Figure \ref{fig:ihc-re-skipping-strategy-varying-threshold}b, Figure \ref{fig:ihc-re-skipping-strategy-varying-threshold}e, and Figure \ref{fig:ihc-re-skipping-strategy-varying-threshold}h), whereas for \ihc such behavior was not observed (Figure \ref{fig:ihc-skipping-strategy-varying-threshold}b, Figure \ref{fig:ihc-skipping-strategy-varying-threshold}e, and Figure \ref{fig:ihc-skipping-strategy-varying-threshold}h).

\section{Discussion}
\label{sec:discussion}

This report has presented our contributions for tackling the problem of probabilistic incremental data cleaning.
In addition to cleaning data in batches, our approach employs mechanisms that keep track of the acquired knowledge over time, embedding such knowledge in our ML models.
To the best of our knowledge, our contributions have led to the first incremental data cleaning framework for performing repairs \textbf{(i)} independently of user interactions, \textbf{(ii)} without requiring prior knowledge about the data such as the number of classes per attribute, and \textbf{(iii)} holistically, being able to simultaneously consider multiple signals for feature generation as well as performing repairs of multiple error types.

Our batchwise approach for data cleaning enabled significant gains in repair quality compared to the evaluated competitors, occasionally outperforming even the HoloClean system, which was our non-incremental baseline approach.
For instance, while HoloClean fixed 22\% of the Soccer-12.5k1pct dataset, the \ihcrei05 approach fixed 57\%.
Overall, \ihcre was the best approach with respect to repair quality, closely followed by the \hcb competitor and by the other variant of our proposal, \ihc.
We managed to use significantly less memory as well, as the proposed approaches required just 10\%--35\% of the maximum amount of memory used by the \hcb competitor (which ends up processing the whole dataset at the last batch).
Furthermore, with respect to execution time, we were overall between \hca and \hcb (our lower- and upper-bound incremental competitors, respectively), but significantly outperforming \hcb while compromising little on repair quality.
Specifically, the \ihc approach with the proposed training skipping mechanism managed to take just 2\%--20\% of the total execution time taken by the \hcb competitor.

\section*{Acknowledgements}

This research was supported by the \textit{Coordena\c{c}\~{a}o de Aperfei\c{c}oamento de Pessoal de N\'{i}vel Superior - Brasil (CAPES) - Finance Code 001}; by the \textit{Funda\c{c}\~{a}o de Amparo \`{a} Pesquisa do Estado de S\~{a}o Paulo (FAPESP)} under Grants \#2015/15392-7, \#2016/17078-0, and \#2018/20360-5; and by the \textit{Conselho Nacional de Desenvolvimento Cient\'{i}fico e Tecnol\'{o}gico (CNPq)}.
The opinions, hypotheses, and conclusions expressed in this work are the authors' responsibility and do not reflect FAPESP's opinion.

\bibliographystyle{acm}
\bibliography{references}

\begin{thebibliography}{10}

\bibitem{Abedjan2016Detecting}
{\sc Abedjan, Z., Chu, X., Deng, D., Fernandez, R.~C., Ilyas, I.~F., Ouzzani,
  M., Papotti, P., Stonebraker, M., and Tang, N.}
\newblock {Detecting Data Errors: Where Are We and What Needs to Be Done?}
\newblock {\em Proceedings of the VLDB Endowment 9}, 12 (Aug. 2016), 993--1004.

\bibitem{Arocena2015BART}
{\sc Arocena, P.~C., Glavic, B., Mecca, G., Miller, R.~J., Papotti, P., and
  Santoro, D.}
\newblock {Messing Up with BART: Error Generation for Evaluating Data-Cleaning
  Algorithms}.
\newblock {\em Proceedings of the VLDB Endowment 9}, 2 (Oct. 2015), 36--47.

\bibitem{Baylor2017TFX}
{\sc Baylor, D., Breck, E., Cheng, H.-T., Fiedel, N., Foo, C.~Y., Haque, Z.,
  Haykal, S., Ispir, M., Jain, V., Koc, L., Koo, C.~Y., Lew, L., Mewald, C.,
  Modi, A.~N., Polyzotis, N., Ramesh, S., Roy, S., Whang, S.~E., Wicke, M.,
  Wilkiewicz, J., Zhang, X., and Zinkevich, M.}
\newblock {TFX: A TensorFlow-Based Production-Scale Machine Learning Platform}.
\newblock In {\em Proceedings of the 23rd ACM SIGKDD International Conference
  on Knowledge Discovery and Data Mining\/} (New York, NY, USA, 2017), KDD `17,
  Association for Computing Machinery, pp.~1387--1395.

\bibitem{Beskales2013OnTheRelative}
{\sc Beskales, G., Ilyas, I.~F., Golab, L., and Galiullin, A.}
\newblock {On the Relative Trust between Inconsistent Data and Inaccurate
  Constraints}.
\newblock In {\em Proceedings of the 29th International Conference on Data
  Engineering\/} (Los Alamitos, CA, USA, 2013), ICDE `13, IEEE Computer
  Society, pp.~541--552.

\bibitem{Chu2013Discovering}
{\sc Chu, X., Ilyas, I.~F., and Papotti, P.}
\newblock {Discovering Denial Constraints}.
\newblock {\em Proceedings of the VLDB Endowment 6}, 13 (Aug. 2013),
  1498--1509.

\bibitem{Heidari2019HoloDetect}
{\sc Heidari, A., McGrath, J., Ilyas, I.~F., and Rekatsinas, T.}
\newblock {HoloDetect: Few-Shot Learning for Error Detection}.
\newblock In {\em Proceedings of the 2019 ACM SIGMOD International Conference
  on Management of Data\/} (New York, NY, USA, 2019), SIGMOD `19, Association
  for Computing Machinery, pp.~829--846.

\bibitem{Ilyas2016Effective}
{\sc Ilyas, I.~F.}
\newblock {Effective Data Cleaning with Continuous Evaluation}.
\newblock {\em IEEE Data Engineering Bulletin 39}, 2 (2016), 38--46.

\bibitem{Ilyas2015Trends}
{\sc Ilyas, I.~F., and Chu, X.}
\newblock {Trends in Cleaning Relational Data: Consistency and Deduplication}.
\newblock {\em Foundations and Trends in Databases 5}, 4 (2015), 281--393.

\bibitem{Ilyas2019Data}
{\sc Ilyas, I.~F., and Chu, X.}
\newblock {\em {Data Cleaning}}.
\newblock Association for Computing Machinery, New York, NY, USA, 2019.

\bibitem{Kolahi2009OnApproximating}
{\sc Kolahi, S., and Lakshmanan, L. V.~S.}
\newblock {On Approximating Optimum Repairs for Functional Dependency
  Violations}.
\newblock In {\em Proceedings of the 12th International Conference on Database
  Theory\/} (New York, NY, USA, 2009), ICDT `09, Association for Computing
  Machinery, pp.~53--62.

\bibitem{Koller2009Probabilistic}
{\sc Koller, D., and Friedman, N.}
\newblock {\em {Probabilistic Graphical Models: Principles and Techniques}}.
\newblock The MIT Press, Cambridge, MA, USA, 2009.

\bibitem{Krishnan2016ActiveClean}
{\sc Krishnan, S., Wang, J., Wu, E., Franklin, M.~J., and Goldberg, K.}
\newblock {ActiveClean: Interactive Data Cleaning for Statistical Modeling}.
\newblock {\em Proceedings of the VLDB Endowment 9}, 12 (Aug. 2016), 948--959.

\bibitem{Mayfield2010ERACER}
{\sc Mayfield, C., Neville, J., and Prabhakar, S.}
\newblock {ERACER: A Database Approach for Statistical Inference and Data
  Cleaning}.
\newblock In {\em Proceedings of the 2010 ACM SIGMOD International Conference
  on Management of Data\/} (New York, NY, USA, 2010), SIGMOD `10, Association
  for Computing Machinery, pp.~75--86.

\bibitem{Naumann2014Data}
{\sc Naumann, F.}
\newblock {Data Profiling Revisited}.
\newblock {\em ACM SIGMOD Record 42}, 4 (Feb. 2014), 40--49.

\bibitem{Papotti2013Holistic}
{\sc Papotti, P., Chu, X., and Ilyas, I.~F.}
\newblock {Holistic Data Cleaning: Putting Violations into Context}.
\newblock In {\em Proceedings of the 29th International Conference on Data
  Engineering\/} (Los Alamitos, CA, USA, 2013), ICDE `13, IEEE Computer
  Society, pp.~458--469.

\bibitem{Rekatsinas2017HoloClean}
{\sc Rekatsinas, T., Chu, X., Ilyas, I.~F., and R{\'e}, C.}
\newblock {HoloClean: Holistic Data Repairs with Probabilistic Inference}.
\newblock {\em Proceedings of the VLDB Endowment 10}, 11 (Aug. 2017),
  1190--1201.

\bibitem{Roh2019Survey}
{\sc Roh, Y., Heo, G., and Whang, S.~E.}
\newblock {A Survey on Data Collection for Machine Learning: A Big Data -- AI
  Integration Perspective}.
\newblock {\em CoRR abs/1811.03402\/} (2018), 1--20.

\bibitem{Schelter2018OnChallenges}
{\sc Schelter, S., Biessmann, F., Januschowski, T., Salinas, D., Seufert, S.,
  and Szarvas, G.}
\newblock {On Challenges in Machine Learning Model Management}.
\newblock {\em Bulletin of the IEEE Computer Society Technical Committee on
  Data Engineering 41}, 4 (2018), 5--13.

\bibitem{Shin2015Incremental}
{\sc Shin, J., Wu, S., Wang, F., De~Sa, C., Zhang, C., and R{\'e}, C.}
\newblock {Incremental Knowledge Base Construction Using DeepDive}.
\newblock {\em Proceedings of the VLDB Endowment 8}, 11 (July 2015),
  1310--1321.

\bibitem{Volkovs2014Continuous}
{\sc Volkovs, M., Chiang, F., Szlichta, J., and Miller, R.~J.}
\newblock {Continuous Data Cleaning}.
\newblock In {\em Proceedings of the 30th International Conference on Data
  Engineering\/} (Los Alamitos, CA, USA, Apr. 2014), vol.~1 of {\em ICDE `14},
  IEEE Computer Society, pp.~244--255.

\bibitem{Yakout2013SCAREd}
{\sc Yakout, M., Berti-{\'E}quille, L., and Elmagarmid, A.~K.}
\newblock {Don't Be SCAREd: Use SCalable Automatic REpairing with Maximal
  Likelihood and Bounded Changes}.
\newblock In {\em Proceedings of the 2013 ACM SIGMOD International Conference
  on Management of Data\/} (New York, NY, USA, 2013), SIGMOD `13, Association
  for Computing Machinery, pp.~553--564.

\bibitem{Zhang2014DimmWitted}
{\sc Zhang, C., and R{\'e}, C.}
\newblock {DimmWitted: A Study of Main-Memory Statistical Analytics}.
\newblock {\em Proceedings of the VLDB Endowment 7}, 12 (Aug. 2014),
  1283--1294.

\end{thebibliography}

\appendix

\section{Incremental Conditional Entropy}
\label{sec:incremental_conditional_entropy}

We formally present how to incrementally maintain the conditional entropy of each pair of attributes in an evolving dataset.

\textbf{Definition 1.}
Let $X$ and $Y$ be two samples.
Define their \textit{conditional entropy} as
\begin{equation}
H(X | Y) = - \sum_{x \in X} \sum_{y \in Y} p(x, y) \log \dfrac{p(x, y)}{p(y)},
\end{equation}
\noindent
where
\begin{itemize}
  \item $p(x,y) = \dfrac{z}{n}$, $n$ being the total number of tuples and $z$ being the number of times values $x \in X$ and $y \in Y$ co-occur,
  \item $p(y) = \dfrac{w}{n}$, $w$ being the number of times value $y \in Y$ occurs.
\end{itemize}
\noindent
Then, the conditional entropy equation can be rewritten as
\begin{equation}
H(X | Y) = - \sum_{i = 1}^{|X|} \sum_{j = 1}^{|Y|} \dfrac{z_{ij}}{n} \log \dfrac{\dfrac{z_{ij}}{n}}{\dfrac{w_j}{n}}.
\end{equation}

\textbf{Lemma 1.}
Let $X$ and $Y$ be two samples and $H_{n}$ be their conditional entropy given $n$ tuples.
For any positive integer $D$,
\begin{equation}
- \sum_{i = 1}^{|X|} \sum_{j = 1}^{|Y|} \dfrac{z_{ij}}{n + D} \log \dfrac{\dfrac{z_{ij}}{n + D}}{\dfrac{w_j}{n + D}} = \dfrac{n}{n + D} \cdot H_{n}.
\end{equation}
\begin{proof}
   Write $\dfrac{z_{ij}}{n + D} = \dfrac{n (z_{ij})}{n (n + D)}$ and $\dfrac{w_j}{n + D} = \dfrac{n (w_j)}{n (n + D)}$. \\[2ex]
   \noindent
   Then, we have
  \begin{align*}
  - \sum_{i = 1}^{|X|} \sum_{j = 1}^{|Y|} \dfrac{z_{ij}}{n + D} \log \dfrac{\dfrac{z_{ij}}{n + D}}{\dfrac{w_j}{n + D}} &= \dfrac{n}{n + D} \left( - \sum_{i = 1}^{|X|} \sum_{j = 1}^{|Y|} \dfrac{z_{ij}}{n} \log \dfrac{\dfrac{z_{ij}}{n}}{\dfrac{w_j}{n}} \right) = \dfrac{n}{n + D} \cdot H_{n}.
  \end{align*}
\end{proof}

The following claim provides an update formula for when a new tuple arrives.

\textbf{Claim 1.}
Let $H_{n}$ be the current samples' conditional entropy and consider an incoming tuple having new attribute values $x' \notin X$ and $y' \notin Y$.
Assume $X'$ and $Y'$ are the up-to-date versions of samples $X$ and $Y$, that is, $X' = X \cup \{x'\}$ and $Y' = Y \cup \{y'\}$.
Then, we have
\begin{equation}
H_{n + 1} = \dfrac{n}{n + 1} \cdot H_{n}.
\end{equation}
\begin{proof}
  By definition we have
  \begin{equation}
  H_{n + 1} = - \sum_{i = 1}^{|X'|} \sum_{j = 1}^{|Y'|} \dfrac{z_{ij}}{n + 1} \log \dfrac{\dfrac{z_{ij}}{n + 1}}{\dfrac{w_j}{n + 1}}.
  \end{equation}
  \noindent
  Considering $x'$ and $y'$ as attribute values in the incoming tuple, this equation can be rewritten as
  \begin{equation*}
  H_{n + 1} = \left( - \sum_{i = 1}^{|X|} \sum_{j = 1}^{|Y|} \dfrac{z_{ij}}{n + 1} \log \dfrac{\dfrac{z_{ij}}{n + 1}}{\dfrac{w_j}{n + 1}} \right) - p(x', y') \log \dfrac{p(x', y')}{p(y')}.
  \end{equation*}
  \noindent
  The new tuple accounts for 1 co-occurrence of values $x' \in X'$ and $y' \in Y'$, hence $p(x', y') = \dfrac{1}{n + 1}$.
  Furthermore, it accounts for 1 occurrence of value $y' \in Y'$, hence $p(y') = \dfrac{1}{n + 1}$.
  So, we have
  \begin{align*}
  H_{n + 1} &= \left( - \sum_{i = 1}^{|X|} \sum_{j = 1}^{|Y|} \dfrac{z_{ij}}{n + 1} \log \dfrac{\dfrac{z_{ij}}{n + 1}}{\dfrac{w_j}{n + 1}} \right) - \dfrac{1}{n + 1} \log \dfrac{\dfrac{1}{n + 1}}{\dfrac{1}{n + 1}} \\[2ex]
  &= - \sum_{i = 1}^{|X|} \sum_{j = 1}^{|Y|} \dfrac{z_{ij}}{n + 1} \log \dfrac{\dfrac{z_{ij}}{n + 1}}{\dfrac{w_j}{n + 1}}.
  \end{align*}
  \noindent
  Therefore, following from Lemma 1, we have
  \begin{equation}
  H_{n + 1} = \dfrac{n}{n + 1} \cdot H_{n}.
  \end{equation}
\end{proof}

The following theorem generalizes Claim 1 by providing a formula for the conditional entropy of concatenated samples, given the conditional entropies $H_{n}$ over $n$ tuples and $G_{m}$ over $m$ tuples.

\textbf{Theorem 1.}
Let $X = \{x_i\}_{i = 1}^{|X|}$, $Y = \{y_i\}_{i = 1}^{|Y|}$, $A = \{a_i\}_{i = 1}^{|A|}$, and $B = \{b_i\}_{i = 1}^{|B|}$ be samples.
Suppose $X$ is to be concatenated with $A$, and $Y$ is to be concatenated with $B$.
Samples $X$ and $A$ can either be disjoint or not, whereas $Y \cap B = \emptyset$.
Assume $n$ is the number of tuples having attribute values from $X$ and $Y$, and $m$ is the number of tuples having attribute values from $A$ and $B$.
Let $H_{n}$ be the conditional entropy of samples $X$ and $Y$, and let $G_{m}$ be the conditional entropy of samples $A$ and $B$.
Finally, let $\hat{X} = X \cup A$ and $\hat{Y} = Y \cup B$ be the concatenated samples.
Define the variables
\begin{itemize}
  \item $z$ as the number of times values $x \in X$ and $y \in Y$ co-occur,
  \item $w$ as the number of times value $y \in Y$ occurs,
  \item $p$ as the number of times values $a \in A$ and $b \in B$ co-occur,
  \item $q$ as the number of times value $b \in B$ occurs,
  \item $\hat{z}$ as the number of times values $\hat{x} \in \hat{X}$ and $\hat{y} \in \hat{Y}$ co-occur,
  \item $\hat{w}$ as the number of times value $\hat{y} \in \hat{Y}$ occurs,
\end{itemize}
\noindent
all of which are greater than $0$.
Then, we have
\begin{equation}
\label{eq:theorem_1}
H_{n + m} = \dfrac{n}{n + m} \cdot H_{n} + \dfrac{m}{n + m} \cdot G_{m}.
\end{equation}
\begin{proof}
  Similarly as before, we have
  \begin{align*}
  H_{n + m} &= - \sum_{i = 1}^{|\hat{X}|} \sum_{j = 1}^{|\hat{Y}|} \dfrac{\hat{z}_{ij}}{n + m} \log \dfrac{\dfrac{\hat{z}_{ij}}{n + m}}{\dfrac{\hat{w}_j}{n + m}} \\[2ex]
  &= - \sum_{i = 1}^{|X|} \sum_{j = 1}^{|Y|} \dfrac{z_{ij}}{n + m} \log \dfrac{\dfrac{z_{ij}}{n + m}}{\dfrac{w_j}{n + m}} - \sum_{i = 1}^{|A|} \sum_{j = 1}^{|B|} \dfrac{p_{ij}}{n + m} \log \dfrac{\dfrac{p_{ij}}{n + m}}{\dfrac{q_j}{n + m}} \\[2ex]
  &= \dfrac{n}{n + m} \cdot H_{n} + \dfrac{m}{n + m} \cdot G_{m},
  \end{align*}
  \noindent
  where the last equality follows from applying Lemma 1 twice.
\end{proof}


The following theorem provides a formula for updating the conditional entropy when a new sample has only unseen attribute values and another new sample has existing attribute values.

\textbf{Theorem 2.}
Let $X$, $Y$, $A$, and $B$ be samples.
Suppose $X$ is to be concatenated with $A$, $Y$ is to be concatenated with $B$, and consider $X \cap A = \emptyset$ and $B \subseteq Y$.
Assume $n$ existing tuples have attribute values from $X$ and $Y$, and $m$ incoming tuples have attribute values from $A$ and $B$.
Let $Y^{+}$ be the attribute values $y^{+} \in Y \cap B$, whose frequencies will have increased after the samples' concatenation, and let $B^{*}$ be the attribute values $b^{*}\in B \setminus Y^{+}$, which are all unseen values to be concatenated to the values $y \in Y$.
Furthermore, let $H_{n}$ be the conditional entropy of $X$ and $Y$ given $n$ tuples.
Define
\begin{itemize}
  \item $z^{+}$ as the number of times values $x \in X$ and $y^{+} \in Y^{+}$ co-occur in the $n$ existing tuples, with $z^{+} > 0$,
  \item $w^{+}$ as the number of times value $y^{+} \in Y^{+}$ occurs in the $n$ existing tuples, with $w^{+} > 0$,
  \item $p^{+}$ as the number of times values $a \in A$ and $y^{+} \in Y^{+}$ co-occur in the $m$ new tuples, with $p^{+} > 0$,
  \item $q^{+}$ as the number of times value $y^{+} \in Y^{+}$ occurs in the $m$ new tuples, with $q^{+} > 0$,
  \item $p^{*}$ as the number of times values $a \in A$ and $b^{*} \in B^{*}$ co-occur in the $m$ new tuples, with $p^{*} \geq 0$,
  \item $q^{*}$ as the number of times value $b^{*} \in B^{*}$ occurs in the $m$ new tuples, with $q^{*} \geq 0$.
\end{itemize}
Then, the samples' conditional entropy $H_{n + m}$ becomes
\begin{equation}
\label{eq:theorem_2}
\begin{split}
H_{n + m} &= \dfrac{n}{n + m} \cdot H_{n} - \left( \sum_{i = 1}^{|X|} \sum_{j = 1}^{|Y^{+}|} \dfrac{z_{ij}^{+}}{n + m} \log \dfrac{\dfrac{z_{ij}^{+}}{n + m}}{\dfrac{w_{j}^{+} + q_{j}^{+}}{n + m}} - \dfrac{z_{ij}^{+}}{n + m} \log \dfrac{\dfrac{z_{ij}^{+}}{n + m}}{\dfrac{w_{j}^{+}}{n + m}} \right) \\[2ex]
&- \left( \sum_{i = 1}^{|A|} \sum_{j = 1}^{|Y^{+}|} \dfrac{p_{ij}^{+}}{n + m} \log \dfrac{\dfrac{p_{ij}^{+}}{n + m}}{\dfrac{w_{j}^{+} + q_{j}^{+}}{n + m}} \right) - \left( \sum_{i = 1}^{|A|} \sum_{j = 1}^{|B^{*}|} \dfrac{p_{ij}^{*}}{n + m} \log \dfrac{\dfrac{p_{ij}^{*}}{n + m}}{\dfrac{q_{j}^{*}}{n + m}} \right).
\end{split}
\end{equation}
\begin{proof}
  We first adjust the conditional entropy value $H_{n}$ to the new number of tuples $n + m$, updating the existing amount of entropy related to values $y^{+} \in Y^{+}$ so that it accounts for their frequencies in the $m$ new tuples as well:
  \begin{align*}
  H_{n}^{*} &= \dfrac{n}{n + m} \cdot H_{n} - \left( \sum_{i = 1}^{|X|} \sum_{j = 1}^{|Y^{+}|} \dfrac{z_{ij}^{+}}{n + m} \log \dfrac{\dfrac{z_{ij}^{+}}{n + m}}{\dfrac{w_{j}^{+} + q_{j}^{+}}{n + m}} - \dfrac{z_{ij}^{+}}{n + m} \log \dfrac{\dfrac{z_{ij}^{+}}{n + m}}{\dfrac{w_{j}^{+}}{n + m}} \right).
  \end{align*}
  \noindent
  Then, taking into account the $m$ new tuples, we have
  \begin{align*}
  H_{n + m} = H_{n}^{*} &- \left( \sum_{i = 1}^{|A|} \sum_{j = 1}^{|Y^{+}|} \dfrac{p_{ij}^{+}}{n + m} \log \dfrac{\dfrac{p_{ij}^{+}}{n + m}}{\dfrac{w_{j}^{+} + q_{j}^{+}}{n + m}} \right) - \left( \sum_{i = 1}^{|A|} \sum_{j = 1}^{|B^{*}|} \dfrac{p_{ij}^{*}}{n + m} \log \dfrac{\dfrac{p_{ij}^{*}}{n + m}}{\dfrac{q_{j}^{*}}{n + m}} \right).
  \end{align*}
  Hence, we have
  \begin{equation*}
  \begin{split}
  H_{n + m} &= \dfrac{n}{n + m} \cdot H_{n} - \left( \sum_{i = 1}^{|X|} \sum_{j = 1}^{|Y^{+}|} \dfrac{z_{ij}^{+}}{n + m} \log \dfrac{\dfrac{z_{ij}^{+}}{n + m}}{\dfrac{w_{j}^{+} + q_{j}^{+}}{n + m}} - \dfrac{z_{ij}^{+}}{n + m} \log \dfrac{\dfrac{z_{ij}^{+}}{n + m}}{\dfrac{w_{j}^{+}}{n + m}} \right) \\[2ex]
  &- \left( \sum_{i = 1}^{|A|} \sum_{j = 1}^{|Y^{+}|} \dfrac{p_{ij}^{+}}{n + m} \log \dfrac{\dfrac{p_{ij}^{+}}{n + m}}{\dfrac{w_{j}^{+} + q_{j}^{+}}{n + m}} \right) - \left( \sum_{i = 1}^{|A|} \sum_{j = 1}^{|B^{*}|} \dfrac{p_{ij}^{*}}{n + m} \log \dfrac{\dfrac{p_{ij}^{*}}{n + m}}{\dfrac{q_{j}^{*}}{n + m}} \right).
  \end{split}
  \end{equation*}
\end{proof}

The next theorem provides a formula for updating the conditional entropy when existing attribute values $x \in X$ and $y \in Y$ have had their frequencies increased, that is, when attribute values from both samples $X$ and $Y$ appear in the incoming tuples.

\textbf{Theorem 3.}
Let $X$ and $Y$ be samples with attribute values from $n$ existing tuples, and let $A$ and $B$ be samples with attribute values from $m$ incoming tuples.
The conditional entropy of $X$ and $Y$ given the $n$ existing tuples is $H_{n}$.
Let $X^{+} \subset X$ and $Y^{+} \subset Y$ be the sets of attribute values that appear both in the $n$ existing tuples and in the $m$ new tuples, and let $X^{=} = X \setminus X^{+}$ be the set of attribute values from $X$ whose frequencies remained the same.
Finally, let $A^{*} = A \setminus X^{+}$ and $B^{*} = B \setminus Y^{+}$ be the sets of unseen attribute values in the $m$ new tuples.
Define the variables
\begin{itemize}
  \item $z^{=+}$ as the number of times values $x^{=} \in X^{=}$ and $y^{+} \in Y^{+}$ co-occur in the $n$ existing tuples,
  \item $z^{++}$ as the number of times values $x^{+} \in X^{+}$ and $y^{+} \in Y^{+}$ co-occur in the $n$ existing tuples,
  \item $p^{++}$ as the number of times values $x^{+} \in X^{+}$ and $y^{+} \in Y^{+}$ co-occur in the $m$ new tuples, with $p^{++} \geq 0$,
  \item $p^{+*}$ as the number of times values $x^{+} \in X^{+}$ and $b^{*} \in B^{*}$ co-occur in the $m$ new tuples, with $p^{+*} > 0$ if $p^{++} = 0$ or $p^{+*} \geq 0$ otherwise,
  \item $r^{*+}$ as the number of times values $a^{*} \in A^{*}$ and $y^{+} \in Y^{+}$ co-occur in the $m$ new tuples, with $r^{*+} > 0$ if $p^{++} = 0$ or $r^{*+} \geq 0$ otherwise,
  \item $r^{**}$ as the number of times values $a^{*} \in A^{*}$ and $b^{*} \in B^{*}$ co-occur in the $m$ new tuples, with $r^{**} = 0$ if $p^{++} = 0$ or $r^{**} \geq 0$ otherwise,
  \item $w^{+}$ as the number of times value $y^{+} \in Y^{+}$ occurs in the $n$ existing tuples,
  \item $q^{+}$ as the number of times value $y^{+} \in Y^{+}$ occurs in the $m$ new tuples,
  \item $s^{*}$ as the number of times value $b^{*} \in B^{*}$ occurs in the $m$ new tuples,
\end{itemize}
\noindent
all of which are greater than $0$, except when specified otherwise.
Then, the conditional entropy $H_{n + m}$ of concatenated samples $\hat{X} = X \cup A$ and $\hat{Y} = Y \cup B$ becomes
\begin{equation}
\begin{split}
H_{n + m} &= \dfrac{n}{n + m} \cdot H_{n} \\[2ex]
&- \left( \sum_{i = 1}^{|X^{=}|} \sum_{j = 1}^{|Y^{+}|} \dfrac{z_{ij}^{=+}}{n + m} \log \dfrac{\dfrac{z_{ij}^{=+}}{n + m}}{\dfrac{w_{j}^{+} + q_{j}^{+}}{n + m}} \right) + \left( \sum_{i = 1}^{|X^{=}|} \sum_{j = 1}^{|Y^{+}|} \dfrac{z_{ij}^{=+}}{n + m} \log \dfrac{\dfrac{z_{ij}^{=+}}{n + m}}{\dfrac{w_{j}^{+}}{n + m}} \right) \\[2ex]
&- \left( \sum_{i = 1}^{|X^{+}|} \sum_{j = 1}^{|Y^{+}|} \dfrac{z_{ij}^{++} + p_{ij}^{++}}{n + m} \log \dfrac{\dfrac{z_{ij}^{++} + p_{ij}^{++}}{n + m}}{\dfrac{w_{j}^{+} + q_{j}^{+}}{n + m}} \right) + \left( \sum_{i = 1}^{|X^{+}|} \sum_{j = 1}^{|Y^{+}|} \dfrac{z_{ij}^{++}}{n + m} \log \dfrac{\dfrac{z_{ij}^{++}}{n + m}}{\dfrac{w_{j}^{+}}{n + m}} \vphantom{\sum_{i = 1}^{|X^{+}|} \sum_{j = 1}^{|Y^{+}|}} \right) \\[2ex]
&- \left( \sum_{i = 1}^{|X^{+}|} \sum_{j = 1}^{|B^{*}|} \dfrac{p_{ij}^{+*}}{n + m} \log \dfrac{\dfrac{p_{ij}^{+*}}{n + m}}{\dfrac{s_{j}^{*}}{n + m}} \right) \\[2ex]
&- \left( \sum_{i = 1}^{|A^{*}|} \sum_{j = 1}^{|Y^{+}|} \dfrac{r_{ij}^{*+}}{n + m} \log \dfrac{\dfrac{r_{ij}^{*+}}{n + m}}{\dfrac{w_{j}^{+} + q_{j}^{+}}{n + m}} \right) - \left( \sum_{i = 1}^{|A^{*}|} \sum_{j = 1}^{|B^{*}|} \dfrac{r_{ij}^{**}}{n + m} \log \dfrac{\dfrac{r_{ij}^{**}}{n + m}}{\dfrac{s_{j}^{*}}{n + m}} \right).
\end{split}
\end{equation}
\begin{proof}
  We begin by adjusting the conditional entropy value $H_{n}$ to the new number of tuples $n + m$, as well as updating the existing amount of entropy of values $y^{+} \in Y^{+}$ whose frequencies increased and that co-occur with values $x^{=} \in X^{=}$:
  \begin{align*}
  H_{n}^{*} &= \dfrac{n}{n + m} \cdot H_{n} - \left( \sum_{i = 1}^{|X^{=}|} \sum_{j = 1}^{|Y^{+}|} \dfrac{z_{ij}^{=+}}{n + m} \log \dfrac{\dfrac{z_{ij}^{=+}}{n + m}}{\dfrac{w_{j}^{+} + q_{j}^{+}}{n + m}} \right) + \left( \sum_{i = 1}^{|X^{=}|} \sum_{j = 1}^{|Y^{+}|} \dfrac{z_{ij}^{=+}}{n + m} \log \dfrac{\dfrac{z_{ij}^{=+}}{n + m}}{\dfrac{w_{j}^{+}}{n + m}} \right).
  \end{align*}
  \noindent
  Then, we update the existing amount of entropy regarding values $x^{+} \in X^{+}$ and $y^{+} \in Y^{+}$:
  \begin{align*}
  H_{n}^{**} &= H_{n}^{*} - \left( \sum_{i = 1}^{|X^{+}|} \sum_{j = 1}^{|Y^{+}|} \dfrac{z_{ij}^{++} + p_{ij}^{++}}{n + m} \log \dfrac{\dfrac{z_{ij}^{++} + p_{ij}^{++}}{n + m}}{\dfrac{w_{j}^{+} + q_{j}^{+}}{n + m}} \right) + \left( \sum_{i = 1}^{|X^{+}|} \sum_{j = 1}^{|Y^{+}|} \dfrac{z_{ij}^{++}}{n + m} \log \dfrac{\dfrac{z_{ij}^{++}}{n + m}}{\dfrac{w_{j}^{+}}{n + m}} \right).
  \end{align*}
  The next step adds the amount of entropy related to values $x^{+} \in X^{+}$ and $b^{*} \in B^{*}$:
  \begin{equation*}
  H_{n}^{***} = H_{n}^{**} - \left( \sum_{i = 1}^{|X^{+}|} \sum_{j = 1}^{|B^{*}|} \dfrac{p_{ij}^{+*}}{n + m} \log \dfrac{\dfrac{p_{ij}^{+*}}{n + m}}{\dfrac{s_{j}^{*}}{n + m}} \right).
  \end{equation*}
  Then, we take into account the unseen values $a^{*} \in A^{*}$.
  We first add the amount of entropy related to values $a^{*} \in A^{*}$ and $y^{+} \in Y^{+}$:
  \begin{equation*}
  H_{n}^{****} = H_{n}^{***} - \left( \sum_{i = 1}^{|A^{*}|} \sum_{j = 1}^{|Y^{+}|} \dfrac{r_{ij}^{*+}}{n + m} \log \dfrac{\dfrac{r_{ij}^{*+}}{n + m}}{\dfrac{w_{j}^{+} + q_{j}^{+}}{n + m}} \right).
  \end{equation*}
  Finally, we add the amount of entropy related to values $a^{*} \in A^{*}$ and $b^{*} \in B^{*}$:
  \begin{equation*}
  H_{n + m} = H_{n}^{****} - \left( \sum_{i = 1}^{|A^{*}|} \sum_{j = 1}^{|B^{*}|} \dfrac{r_{ij}^{**}}{n + m} \log \dfrac{\dfrac{r_{ij}^{**}}{n + m}}{\dfrac{s_{j}^{*}}{n + m}} \right).
  \end{equation*}
\end{proof}


\end{document}